\documentclass[preprint]{aastex631}
\begin{document}

\title{FAINT WHITE DWARF FLUX STANDARDS: DATA AND MODELS}

\author{Ralph~C.\ Bohlin} \affiliation{Space Telescope Science Institute, 3700
	San Martin Drive, Baltimore,  MD 21218, USA}
\author{Susana Deustua} \affiliation{Sensor Science Division, National Institute
	of Standards and Technology, Gaithersburg, MD 20899-8441, USA}
\author{Gautham Narayan} \affiliation{University of Illinois at
	Urbana-Champaign, 1002 W. Green Street, Urbana, IL 61801, USA}
\author{Abhijit Saha} \affiliation{NSF's National Optical Infrared Astronomy
	Research Laboratory, 950 North Cherry Avenue, Tucson, AZ 85719, USA}
\author{Annalisa Calamida} \affiliation{Space Telescope Science Institute, 3700
	San Martin Drive, Baltimore,  MD 21218, USA}
\author{Karl D. Gordon} \affiliation{Space Telescope Science Institute, 3700 San
	Martin Drive, Baltimore,  MD 21218, USA}
\author{Jay B. Holberg} \affiliation{The University of Arizona, Lunar and Planetary
	Laboratory, 1629 East University Boulevard, Tucson, AZ 85721, USA}
\author{Ivan Hubeny} \affiliation{The University of Arizona, Steward Observatory,
	933 North Cherry Avenue, Tucson, AZ 85721, USA}
\author{Thomas Matheson} \affiliation{NSF's National Optical Infrared Astronomy
	Research Laboratory, 950 North Cherry Avenue, Tucson, AZ 85719, USA}
\author{Armin Rest} \affiliation{Space Telescope Science Institute, 3700 San
	Martin  Drive, Baltimore, MD 21218, USA and Department of Physics and
	Astronomy, Johns Hopkins University, Baltimore, MD 21218, USA}

\begin{abstract}

Fainter standard stars are essential for the calibration of larger telescopes. 
This work adds to the CALSPEC (calibration spectra) database 19 faint
white dwarfs (WDs) with all-sky  coverage and V magnitudes between 16.5 and
18.7. Included for these stars is new UV (ultraviolet) HST
(Hubble Space Telescope) STIS (Space Telescope Imaging
Spectrometer) spectrophotometry between 1150 and 3000~\AA\ with a resolution of
$\sim$500. Pure hydrogen WD models are fit to these UV spectra and to six-band
HST/WFC3 (Wide Field Camera 3) photometry at 0.28 to 1.6~\micron\ to
construct predicted model SEDs (spectral energy distributions) covering
wavelengths from 900~\AA\ to the JWST (James Webb Space Telescope)
limit of 30~\micron\ using well-established CALSPEC procedures for producing
flux standards with the goal of 1\% accuracy.

\end{abstract}


\section{Introduction}			

\citet{bohlin2020} and references therein discuss the
CALSPEC\footnote{http://www.stsci.edu/hst/instrumentation/reference-data-for-calibration-and-tools/astronomical-catalogs/calspec}
database of  SEDs for stellar flux standards. The CALSPEC SEDs are based on
HST/STIS flux calibrated spectra, sometimes supplemented by HST/WFC3 and  HST
NICMOS (Near Infrared Camera and Multi-Object Spectrometer)
spectrophotometry. Model atmosphere grids of stellar spectra are fit to these
measured flux distributions to estimate the flux at longer wavelengths than
covered by the HST observations. For this work, the Hubeny grid version 207 is
used to fit the STIS SEDs. \citet{bohlin2020} introduced this
grid\footnote{DOI10.17909/t9-7myn-4y46}, which contains 132 models with
effective temperature ($T_\mathrm{eff}$) in the range 20,000-95,000~K and
surface gravity ($\log g$) between 7.0 and 9.5, with six steps of 0.5, where g
has units of cm~s$^{-2}$. The steps in $T_\mathrm{eff}$ are 2,000~K between
20,000 and 40,000~K and 5,000~K between 40,000 and 95,000~K. 

The normal CALSPEC SED includes a mix of the observed spectrophotometry with
model extrapolations to longer wavelengths. The fitted models provide standard
star flux distributions for calibration of IR (infrared)
instrumentation. Most of the current JWST CALSPEC standards are too bright (V
$<$16) for many of the standard detector modes and provide only indirect flux
calibrations via subarray data. However, our new fainter stars (V=16 to 19) can
be observed in the standard science modes, thus avoiding any uncertainty
associated with the small detector subarray modes used for bright stars. The
JWST flux calibration plan \citep{gordon2022} utilizes three categories of
CALSPEC standards: hot, A type, and G type stars, while WDs fall in the hot star
category.

Our long term project to establish a network of faint WD stars with complete sky
coverage provides an ideal set of fainter IR flux standards for JWST. These
faint star SEDs were based on HST/WFC3 photometry in six filters for 35 WDs from
the A. Saha programs 12967, 13711, and 15113. \citet{narayan2019} published
preliminary results, while \citet{calam2022perf} present a variability analysis
and finding charts. Details of the WFC3 photometric data reduction are in
\citet{calamida2019}, which includes a description of the ILAPH photometry.
\citet{axelrod2023} fit this six-filter WFC3 photometry to
sub-percent precision with pure-hydrogen model atmosphere SEDs using a
hierarchical Bayesian analysis process similar to the preliminary analysis of
\citet{narayan2019}.

The \citet{axelrod2023} paper examines the internal consistency of the SEDs of
all 35 DA white dwarfs that span a wide range of temperatures and surface
gravities and includes the three primary WDs that are used to define the flux
scale for CALSPEC. This analysis compares their measured fluxes from the near-UV
to the near-IR against predictions from the \citet{Hubeny1995} NLTE
(non-local thermodynamic equilibrium) models of DA white dwarf
atmospheres and results in a rigid lattice of SEDs on a physical basis and in
relative apparent brightness for the set of stars that is self-consistent to a
few milli-mag rms within this wavelength domain. The method of analysis is
independent of CALSPEC, except for effectively borrowing the zero-point used in
the 2014 version of the CALSPEC (CALSPEC14) flux scale for the absolute flux of
Vega at 5556~\AA\ (air) to achromatically tie this 'lattice' to an absolute flux
scale. While CALSPEC14 and \citet{axelrod2023} both used the same three primary
DA white dwarf model SEDs, the current implementation of CALSPEC differs in two
respects: a) it is based on newer models, which extend the wavelength coverage
to the JWST limit of 30~\micron; and b) its absolute fluxes reconcile absolute
flux measures of Sirius in the midIR with the Vega 5556~\AA\ flux to determine
the achromatic flux zero-point. The analysis in this paper is tied to the
current \citet{bohlin2020} CALSPEC flux scale and, thus, can differ
systematically from that derived in \citet{axelrod2023}. In addition, the fitted
data sets and methods of analyses differ.  

To supplement the WFC3 photometric constraints on the model fits, the HST
program 16764 (G. Narayan PI) obtained shorter wavelength coverage with the STIS
G140L and G230L gratings for 19 of the 32 stars in \citet{axelrod2023} that are
bright enough to obtain STIS UV spectrophotometry between 1150 and 3000~\AA. Our
goal is to produce CALSPEC SEDs extending to the JWST limit of 30~\micron\ by
fitting both the original WFC3 photometry along with the new STIS
spectrophotometry. Consequent to the statements in the above paragraph, minor
updates bring the Axelrod photometry onto the current CALSPEC flux scale system;
and more recent NLTE models extend the wavelength coverage to 30~\micron. 

An analysis with the standard $\chi^2$ technique of \citet{bohlin2020} used for
all the current CALSPEC models provides consistent SEDs for on-going
JWST flux calibrations with fainter standards than are currently available. JWST
results will provide feedback to our continuing analyses that update these
preliminary results. The definition of flux standards is never finalized
but is a continually evolving process of improvement as the data
set and analysis techniques mature.

Section 2 describes the HST data, while Section 3 explains the procedure of
finding the models that best fit the data. Section 4 
summarizes the results and the future plans for improvements.

\section{HST Data REDUCTION}				

\subsection{STIS}				


The STIS spectra
are extracted from the raw STIS images using a custom set of software provided
by the original STIS Instrument Definition Team and Don Lindler. The advantages
of custom software over the STScI pipeline products include automatic repair of
hot pixels, optimization of cosmic ray rejection, editing of residual noise
spikes, wider extraction heights to improve CCD photometric precision,
co-addition of multiple observations, merging of the separate grating spectra
into one complete SED, and correction of any small offsets from the wavelengths
defined by the STIS emission line calibration spectra. In addition, every new
monitor observation (about three times per year) of the MAMA standard GRW+70
$^{\circ}$5824 or the CCD standard AGK+81$^{\circ}$266 results in a complete
updated flux calibration \citep{bohlin2020} that accounts for the changes in
instrumental sensitivity and is applicable to all of the CALSPEC stars. IDL and
Python versions of these STIS analysis routines reside in 
Github\footnote{https://github.com/spacetelescope/ABSCAL}.

\subsection{WFC3}				

In addition to the UV STIS spectrophotometry, the WFC3 photometry in six filters
constrains the model fits over the wavelength range from the near-UV to the IR
filter F160W at a pivot wavelength of 15369~\AA. The \citet{axelrod2023} 32 SEDs
have absolute fluxes defined by the three CALSPEC *mod\_010.fits SEDs for
the three primary WDs, G191B2B, GD153, and GD71, on the flux scale of
\citet{bohlinetal14}, which was current for the original analysis of
\citet{narayan2019}, while the present HST flux scale is defined by the three
*mod\_012.fits SEDs of \citet{bohlin2020}. These three CALSPEC SEDs for the
primary WDs provide the defining basis for all of the HST flux calibrations,
including both STIS spectrophotometry and WFC3 photometry.

Thus, the transformation of the \citet{axelrod2023} 2014 absolute photometry to
the current 2020 flux scale is just the average ratios of the synthetic
photometry for the three 2020 *mod\_012.fits SEDs to the synthetic photometry of
the 2014 *mod\_010.fits SEDs, as presented in Table~\ref{table:synprim}. As
explained in detail in \citet{bohlin2020}, the transformation from the 2014 to
the 2020 flux has an absolute gray flux increase of 0.87\% and a wavelength 
dependent change due to the improved models for the three stars. The combination
of these two updates total to $<$2\% and bring the \citet{axelrod2023}
photometry from the 2014 to the current 2020 HST photometric scale of
Table~\ref{table:2020}. \citet{axelrod2023} provides both the original WFC3
photometry and a corrected version updated for any WFC3 flux calibration errors,
such as poor bandpass transmission functions. To use the corrected values, the
bandpass transmission must also be corrected to maintain consistency between the
corrected photometry and the synthetic photometry. Here, the analysis begins
with the original photometry, in order to avoid questions of the true bandpass 
and to compare any systematic WFC3 photometry residuals with those same
residuals found by Axelrod.
\begin{deluxetable}{ccccc}     
\tablewidth{0pt}
\tablecolumns{5}
\tablecaption{\label{table:synprim} \mbox{2020 vs. 2014 Effective Flux}}
\tablehead{
\colhead{Star} &\colhead{Filter} &\colhead{2014\tablenotemark{a}} 
	&\colhead{2020\tablenotemark{a}} &\colhead{Ratio} }
\startdata
G191B2B   &F275W &9.4780e-13 &9.6486e-13  &1.0180\\
  GD153   &F275W &1.9550e-13 &1.9967e-13  &1.0213\\
   GD71   &F275W &2.3789e-13 &2.4361e-13  &1.0240\\
AVERAGE   &	 &           &            &1.02112\\
\\
G191B2B   &F336W &4.2643e-13 &4.3315e-13  &1.0158\\
  GD153   &F336W &9.0853e-14 &9.2391e-14  &1.0169\\
   GD71   &F336W &1.1254e-13 &1.1475e-13  &1.0196\\
AVERAGE   &	 &           &            &1.01742\\
\\
G191B2B   &F475W &1.1991e-13 &1.2133e-13  &1.0119\\
  GD153   &F475W &2.7642e-14 &2.7878e-14  &1.0085\\
   GD71   &F475W &3.6412e-14 &3.6824e-14  &1.0113\\
AVERAGE   &	 &           &            &1.01058\\
\\
G191B2B   &F625W &4.3035e-14 &4.3423e-14  &1.0090\\
  GD153   &F625W &1.0194e-14 &1.0245e-14  &1.0050\\
   GD71   &F625W &1.3687e-14 &1.3785e-14  &1.0072\\
AVERAGE   &	 &           &            &1.00708\\
\\
G191B2B   &F775W &1.9553e-14 &1.9680e-14  &1.0065\\
  GD153   &F775W &4.6757e-15 &4.6946e-15  &1.0040\\
   GD71   &F775W &6.3272e-15 &6.3650e-15  &1.0060\\
AVERAGE   &      &           &            &1.00551\\
\\
G191B2B   &F160W &1.2958e-15 &1.2935e-15  &0.9982\\
  GD153   &F160W &3.1516e-16 &3.1571e-16  &1.0017\\
   GD71   &F160W &4.3410e-16 &4.3540e-16  &1.0030\\
AVERAGE   &      &           &            &1.00097\\
\enddata
\tablenotetext{a}{erg~s$^{-1}$ cm$^{-2}$ \AA$^{-1}$}
\tablecomments{Synthetic absolute effective fluxes for the three primary WD
stars in six WFC3 filters as computed from their 2014 CALSPEC SEDs
(*mod\_010.fits) and the 2020 SEDs (*mod\_012.fits). The average ratio of the
2020 to the 2014 fluxes defines the six updates required to convert all of the
\citet{axelrod2023} photometry to the current 2020 CALSPEC flux scale.}
\end{deluxetable}

The STIS G230L grating covers the same wavelengths as the WFC3 F275W filter; and
the synthetic STIS F275W photometry is close to the original WFC3 photometry,
which testifies to the precision of our WFC3 effective fluxes. On average, the
original Axelrod WFC3 photometry is 0.003 mag fainter than the STIS synthetic
photometry; and the rms difference between the synthetic and actual photometry
is 0.006 mag for our 19 WDs, which is consistent  with the STIS G230L broadband
repeatability of 0.002--0.003 mag \citep{bohlin2019} for brighter stars.

\begin{deluxetable}{ccccccc}     
\tablewidth{0pt}
\tablecolumns{7}
\tablecaption{\label{table:2020} \mbox{2020 WFC3 AB magnitudes}}
\tablehead{
\colhead{Star} &\colhead{F275W} &\colhead{F336W} 
	&\colhead{F475W} &\colhead{F625W} &\colhead{ F775W} &\colhead{ F160W}}
\startdata
G191B2B    &  10.467& 10.871& 11.488& 12.023& 12.445& 13.884\\
GD153	   &  12.179& 12.549& 13.089& 13.590& 13.996& 15.413\\
GD71	   &  11.966& 12.317& 12.788& 13.271& 13.666& 15.067\\
WDFS0103-00&  18.172& 18.508& 19.072& 19.561& 19.959& 21.354\\
WDFS0122-30&  17.648& 17.975& 18.449& 18.914& 19.314& 20.704\\
WDFS0228-08&  19.495& 19.696& 19.804& 20.161& 20.495& 21.736\\
WDFS0238-36&  17.767& 17.953& 18.084& 18.431& 18.751& 19.991\\
WDFS0248+33&  17.806& 18.021& 18.359& 18.738& 19.071& 20.339\\
WDFS0458-56&  17.000& 17.332& 17.743& 18.209& 18.595& 19.998\\
WDFS0541-19&  17.998& 18.196& 18.265& 18.616& 18.954& 20.193\\
WDFS0639-57&  17.299& 17.634& 18.167& 18.631& 19.011& 20.379\\
WDFS0727+32&  17.141& 17.452& 17.982& 18.449& 18.831& 20.216\\
WDFS0815+07&  18.927& 19.245& 19.705& 20.176& 20.573& 21.961\\
WDFS0956-38&  17.675& 17.840& 17.851& 18.171& 18.491& 19.689\\
WDFS1024-00&  18.238& 18.495& 18.893& 19.309& 19.659& 20.990\\
WDFS1055-36&  17.347& 17.634& 18.002& 18.419& 18.787& 20.134\\
WDFS1110-17&  17.018& 17.335& 17.856& 18.306& 18.683& 20.056\\
WDFS1111+39&  17.420& 17.811& 18.410& 18.931& 19.338& 20.796\\
WDFS1206+02&  18.217& 18.470& 18.661& 19.052& 19.405& 20.702\\
WDFS1206-27&  15.714& 16.022& 16.465& 16.915& 17.287& 18.648\\
WDFS1214+45&  16.917& 17.264& 17.750& 18.228& 18.623& 20.037\\
WDFS1302+10&  16.165& 16.503& 17.025& 17.506& 17.898& 19.302\\
WDFS1314-03&  18.235& 18.578& 19.091& 19.559& 19.949& 21.327\\
WDFS1434-28&  17.815& 17.958& 17.957& 18.277& 18.578& 19.758\\
WDFS1514+00&  15.087& 15.372& 15.698& 16.112& 16.465& 17.786\\
WDFS1535-77&  15.576& 15.950& 16.542& 17.042& 17.451& 18.889\\
WDFS1557+55&  16.477& 16.858& 17.459& 17.984& 18.382& 19.833\\
WDFS1638+00&  17.993& 18.299& 18.829& 19.273& 19.654& 20.995\\
WDFS1814+78&  15.768& 16.102& 16.533& 16.998& 17.387& 18.785\\
WDFS1837-70&  17.619& 17.772& 17.759& 18.084& 18.405& 19.605\\
WDFS1930-52&  16.706& 17.015& 17.473& 17.919& 18.295& 19.654\\
WDFS2101-05&  18.045& 18.315& 18.645& 19.056& 19.408& 20.739\\
WDFS2317-29&  17.874& 18.122& 18.338& 18.740& 19.100& 20.422\\
WDFS2329+00&  17.920& 18.090& 18.150& 18.462& 18.769& 19.994\\
WDFS2351+37&  17.426& 17.643& 18.064& 18.451& 18.781& 20.074\\
\enddata
\tablecomments{WFC3 photometry on the 2020 CALSPEC flux scale. The
Table~\ref{table:synprim} corrections convert the original WFC3 photometry
from the table 1 \citet{axelrod2023} 2014 flux scale to the current 2020 scale.}
\end{deluxetable}

\section{Result from Fitting Models}			

To extend the flux distributions to longer and shorter wavelengths, model
atmosphere SEDs are fit to the STIS and WFC3 data with the reduced $\chi^2$
technique and the Hubeny \textsc{tlusty} pure hydrogen models in the standard
CALSPEC style \citep{bohlin2020}. To find the best fit model, the WFC3
photometry is compared to synthetic photometry for the model SED; and the STIS
data are fit in bins that avoid the often strong interstellar lines and regions
with weak signal and high photon noise. Table~\ref{table:stisbins} lists these
wavelength bins. The model fitting finds the smallest $\chi^2$
\citep{bohlin2017, bohlinsed2019, bohlin2020} by varying $T_\mathrm{eff}$ and
E(B-V), which are the model effective temperature and selective extinction.

\begin{deluxetable}{c} 			
\tablewidth{0pt}
\tablecaption{\label{table:stisbins} \mbox{Broad Bands for Fitting WD Models}}
\tablehead{
\colhead{Wavelength Ranges (\AA)}}
\startdata 1350--1500\tablenotemark{a} 1500--1600 2000--2100\tablenotemark{b} 
 2100--2200\tablenotemark{b} 2200--2300\tablenotemark{b}
 2300--2400\tablenotemark{b} 2400--2500\tablenotemark{b} 
 2500--2800\tablenotemark{c}\\
\enddata
\tablenotetext{a} {\mbox{Not used for four of the coolest stars WDFS0956-38,
WDFS1434-28,} \mbox{WDFS1837-70, and  WDFS2317-29 because of confusion with the}
\mbox{strong quasi-molecular hydrogen feature near 1400~\AA} in the Hubeny
models.}
\tablenotetext{b} {\mbox{Not used for WDFS1514+00 and WDFS2351+37, because the
2175~\AA} \mbox{extinction bump is anomalously strong and cannot be fit with any
of} \mbox{the \citet{gordon2023} average extinction curves}}
\tablenotemark{c}{\mbox{Replaced by 2500--3100 for WDFS1514+00 and}
\mbox{WDFS2351+37 that have very strong extinction}}
\end{deluxetable}

\subsection{Details of the $\chi^2$ Fitting}		

\begin{figure}			
\centering 
\includegraphics*[width=.9\textwidth,trim=40 60 35 40]{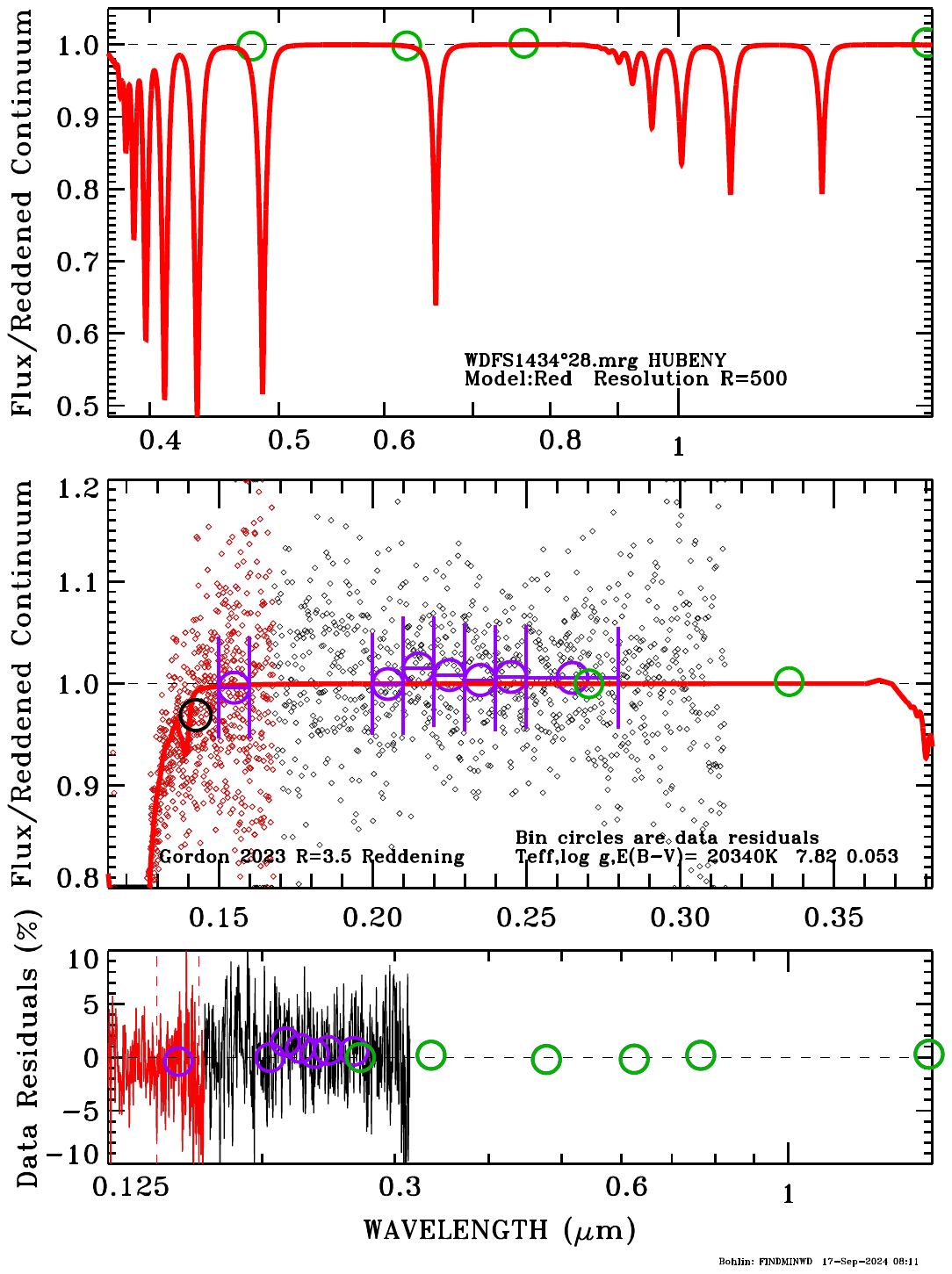}
\caption{\baselineskip=12pt
Model fit and residuals for WDFS1434-28 with the parameters from
Table~\ref{table:modpar} written on the plot. The heavy red line in the top two
panels represents the model divided by its continuum to flatten the display
and preserve spectral features, while in all three panels the
residuals are large green circles for the WFC3 photometry, large purple
circles for the useful STIS bins, and a large black
circle for the STIS G140L bin that is not used as a constraint to the fit.
Middle panel:
Small diamonds are STIS data for G140 (red) and G230L (black). The dip in the
model near 1400~\AA\ is a quasi-molecular hydrogen feature.
Bottom panel:
The purple bars that indicate the bin width and the black circle are omitted
for clarity. The red and black STIS data are binned to a resolution R = 500. 
\label{findmin1434}} \end{figure}

The models from the Hubeny \textsc{tlusty} grid are compared to the STIS
observations in the wavelength bins in Table~\ref{table:stisbins}, and synthetic
photometry of the models is compared to the actual WFC3 photometry using error
estimates for STIS, WFC3, and the models to form the $\chi^2$ sums. Modeling
uncertainties are proportional to the amount of line-blanketing in each STIS bin
or in each WFC3 bandpass. The best fit is found by varying the $T_\mathrm{eff}$
and E(B-V) to find the minimum of the reduced $\chi^2$ value that defines the
model. The mesh size of the model search parameters are 10~K in $T_\mathrm{eff}$
and 0.001 in E(B-V). Our analysis does not consider the constraints of
the ground-based Balmer spectra on the surface gravity ($\log g$), while Axelrod
does utilize the  Balmer lines. The pressure  broadened Balmer line profiles are
most sensitive to $\log g$; and therefore, the Axelrod $\log g$ values are
used and not varied in the fitting. An example of the model fitting appears in
Figure~\ref{findmin1434} for WDFS1434-28 with the noisy STIS G140L data shown in
red in the bottom two panels.

The $\chi^2$ technique relies on minimizing the sum of the squares of the ratio
of the residual differences of the data from the model, as divided by the
uncertainty estimates. The wide STIS bands used for fitting the model minimize
the uncertainty from  photon noise. The WFC3 rms photometric repeatabilities are
0.3\% for the five UVIS filters and 0.6\% for F160W \citep{axelrod2023}. 
For STIS, the repeatability as a function of
wavelength for the monitor star GRW+70 $^{\circ}$5824 is of order 0.7\% for
G140L and 0.3\% for G230L \citep{bohlin2019}. In addition, the STIS background
uncertainty is 4\% of the sky background level to account for the large
separation of the  background regions from the spectral trace. The two
uncertainties are combined in quadrature for each STIS bin; and the relative
weights of each STIS bin and WFC3 photometry point are their individual $\chi^2$
values, which are then all combined in quadrature to get the total $\chi^2$. The
minimum total $\chi^2$ defines the $T_\mathrm{eff}$ and E(B-V) of the
best fitting SED for each WD. To ensure finding the global minimum and not a
local minimum $\chi^2$, the process is iterated to convergence with
perturbations of 30K in $T_\mathrm{eff}$.

\subsection{Extinction by Interstellar Dust}		

The proper accounting for extinction from interstellar dust is crucial to
precisely fitting measured SEDs to theoretical models, which are unextinguished.
If A(V) is the total extinction in magnitude units from interstellar dust in the
V band and the selective extinction E(B-V) is defined as A(B)-A(V) where A(B) is
the extinction in the B band, then average extinction curves are specified in
terms of the parameter R(V)=A(V)/E(B-V). Here, the default is R(V)=3.1, which is
the average over the sky that is also adopted by \citet{axelrod2023}. The
average extinction prescriptions for R(V) are from \citet{gordon2023}, but there
are frequent deviations from these averages for individual sightlines 
\citep{witt1984, mathis1992, valencic2004,gordon2023}.

The larger the
extinction A(V), the larger the effect on the observed SED; and the extinction
is somewhat degenerate with $T_\mathrm{eff}$ in the sense that larger A(V) and
lower $T_\mathrm{eff}$ both reduce the UV flux with respect to longer
wavelengths. The extended wavelength coverage with the new STIS data reduces the
degeneracy in comparison to the smaller wavelength range of just the WFC3 data
alone.

For four stars with some of the higher extinctions in Table~\ref{table:modpar},
exceptions to R(V)=3.1 are required, in order to bring the residuals to
$\lessapprox1\%$ for all STIS bins and all WFC3 photometry. Because these larger
values of A(V) have the biggest effect on the observed SED, deviations from the
default R(V)=3.1 can significantly change the quality of the model fits. For
example for WDFS0639-57, the worst WFC3 residual is for F160W, which drops from
2.8\% to 0.3\%; and the overall $\chi^2$ of the fit decreases from 7.7 to 2.0
for the fits with R(V)=3.1 vs. R(V)=4.2.

For many stars, fitting the G140L data is not possible, even if the restriction
of R(V)=3.1 is relaxed. Thus, the model fitting for these 12 stars proceeds by
omitting the two G140L bins. Because the G140L flux is mostly lower than the
model fit to the longer wavelength data, random photometric repeatabilty for our
faint stars seems an unlikely culprit. For the 12 stars where G140L is not a
constraint, the average G140L residuals in the final column of
Table~\ref{table:modpar} are often many sigma when compared to the $\approx1\%$
repeatability for the STIS monitor star GRW+70 $^{\circ}$5824. These residuals
are with respect to the model fitted to the remaining STIS and WFC3 data after
omission of the two G140L bins. Other possibilities for the poor fits to G140L
include (a) faint cooler sources within $\sim$0.5\arcsec\ that contribute a few
percent to the longer wavelengths but not to the FUV G140L signal or (b) heavy
FUV metal line-blanketing, as is the case for G191B2B, but which has only
$\approx1\%$ FUV blanketing.

\begin{deluxetable}{lccccc}     
\tablewidth{0pt}
\tablecolumns{6}
\tablecaption{\label{table:modpar} \mbox{Model parameters}}
\tablehead{
\colhead{Star} &\colhead{Teff (K)} &\colhead{log g} &\colhead{A(V)}
	&\colhead{R(V)} &\colhead{G140L Resid (\%)} }
\startdata
WDFS0122-30  &34150 &7.77 &0.043  & 3.1 &-4.7  \\
WDFS0248+33  &33230 &7.10 &0.323  & 3.4 & ...  \\
WDFS0458-56  &30290 &7.79 &0.009  & 3.1 & ...  \\
WDFS0639-57  &53230 &7.90 &0.189  & 4.2 &-4.5  \\
WDFS0956-38  &20100 &7.88 &0.112  & 3.1 & ...  \\
WDFS1055-36  &29570 &7.93 &0.093  & 3.1 &-3.6  \\
WDFS1110-17  &46700 &8.01 &0.146  & 3.1 &-4.2  \\
WDFS1206-27  &33980 &7.90 &0.099  & 3.1 &-1.6  \\
WDFS1214+45  &34170 &7.85 &0.003  & 3.1 &-2.0  \\
WDFS1302+10  &42090 &7.93 &0.074  & 3.1 &-7.8  \\
WDFS1434-28  &20340 &7.82 &0.185  & 3.5 & ...  \\
WDFS1514+00  &29200 &7.90 &0.121  & 3.1 &+4.3  \\
WDFS1535-77  &49940 &9.08 &0.012  & 3.1 &-2.7  \\
WDFS1557+55  &57830 &7.55 &0.012  & 3.1 &-2.3  \\
WDFS1814+78  &31040 &7.80 &0.006  & 3.1 &-2.3  \\
WDFS1837-70  &19810 &7.87 &0.105  & 3.1 & ...  \\
WDFS1930-52  &36300 &7.67 &0.133  & 3.5 & ...  \\
WDFS2317-29  &24030 &7.85 &0.022  & 3.1 & ...  \\
WDFS2351+37  &41290 &7.70 &0.313  & 3.1 &+5.4  \\
\enddata
\end{deluxetable}

\begin{figure}			
\centering 
\includegraphics*[width=.9\textwidth,trim=40 60 35 40]{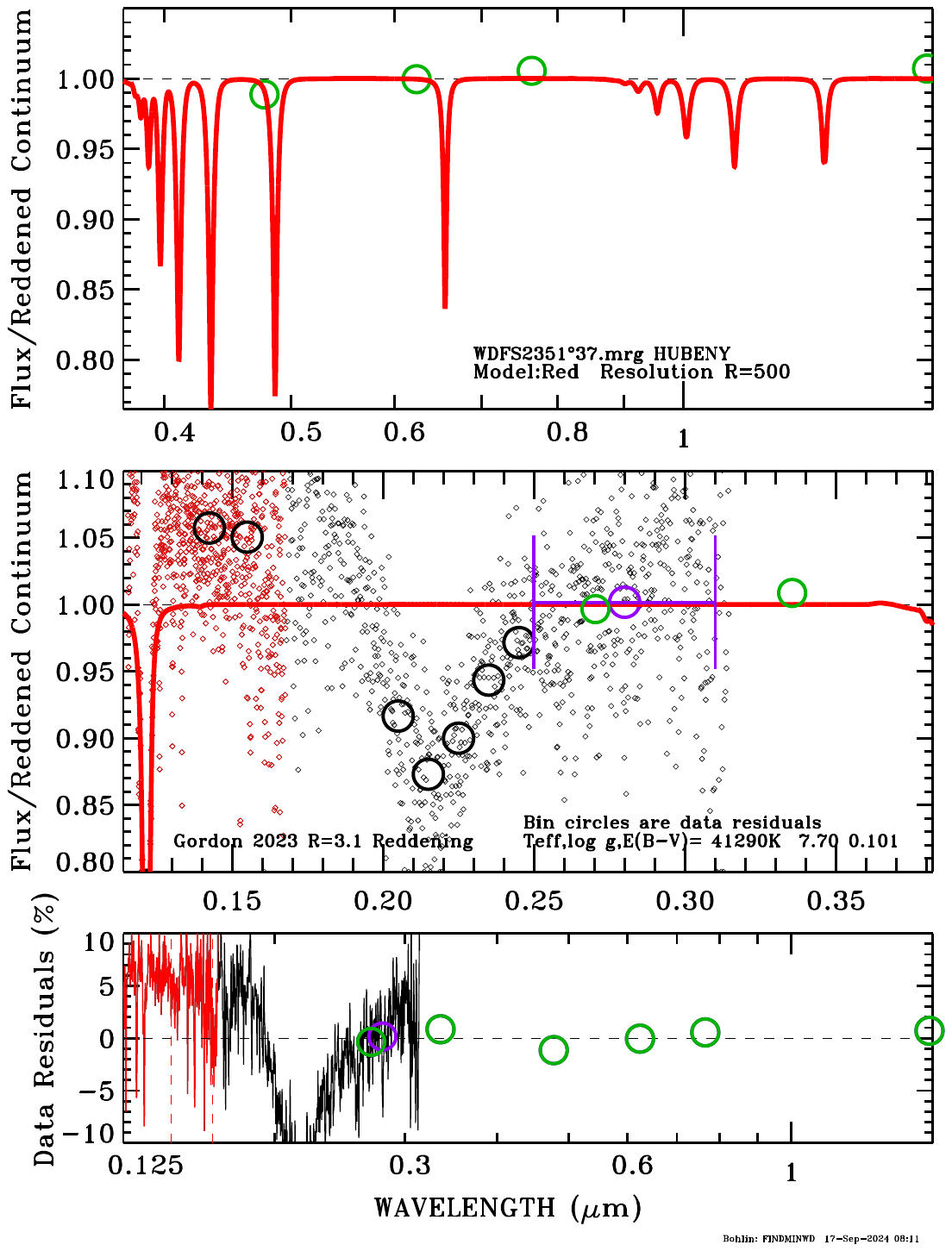}
\caption{\baselineskip=12pt
Model fit and residuals for WDFS2351+37 as in Figure~\ref{findmin1434}. Seven
STIS bins (large black circles) are omitted as constraints to the model
fitting. The bottom panel shows the large STIS spectral residuals and the small
residuals for the WFC3 photometry (green circles).  The full figure set
for the 17 WDs, beside Figure~\ref{findmin1434} and
Figure~\ref{findmin2351}, is available in the electronic version of this paper.
[editor: please put the link to the Fig set here.]
\label{findmin2351}}
\end{figure}

Perhaps, the most likely reason for poor fits is that the reddening is anomalous
and does not correspond to any average extinction curve. \citet{gordon2023} show
considerable scatter in A($\lambda$)/A(V) at $\lambda$=1500~\AA\ and R(V) near
3.1. Figure~\ref{findmin2351} illustrates the  worst case residuals for
WDFS2351+37 that cannot be fit by any \citet{gordon2023} average extinction. The
fit is only to the data longward of 2500~\AA, and the omitted short wavelength
STIS bins appear as open black circles. In the middle and lower panels, the red
model lies above the flux in the 2175~\AA\ extinction bump and below the G140L
STIS data, which suggests that the extinction is anomalously low in the
1500~\AA\ region compared to a stronger than average 2175~\AA\ extinction bump.
Similarly, WDFS1514+00 also has a positive G140L residual in
Table~\ref{table:modpar}. No average dust extinction curve of \citet{gordon2023}
provides a good fit to the observed ultraviolet SED of these two stars.

\textit{The most reliable model flux extrapolations in the IR to 32~\micron\ are
for the seven stars with models that do fit the G140L STIS data and, thus, have
no entry in the G140L Residual column of Table~\ref{table:modpar}.}

\figsetstart
\figsetnum{3}
\figsettitle{Model fit and residuals for the other 17} WDs

\figsetgrpstart
\figsetgrpnum{figurenumber.1}
\figsetgrptitle{WDFS0122-30}
\figsetplot{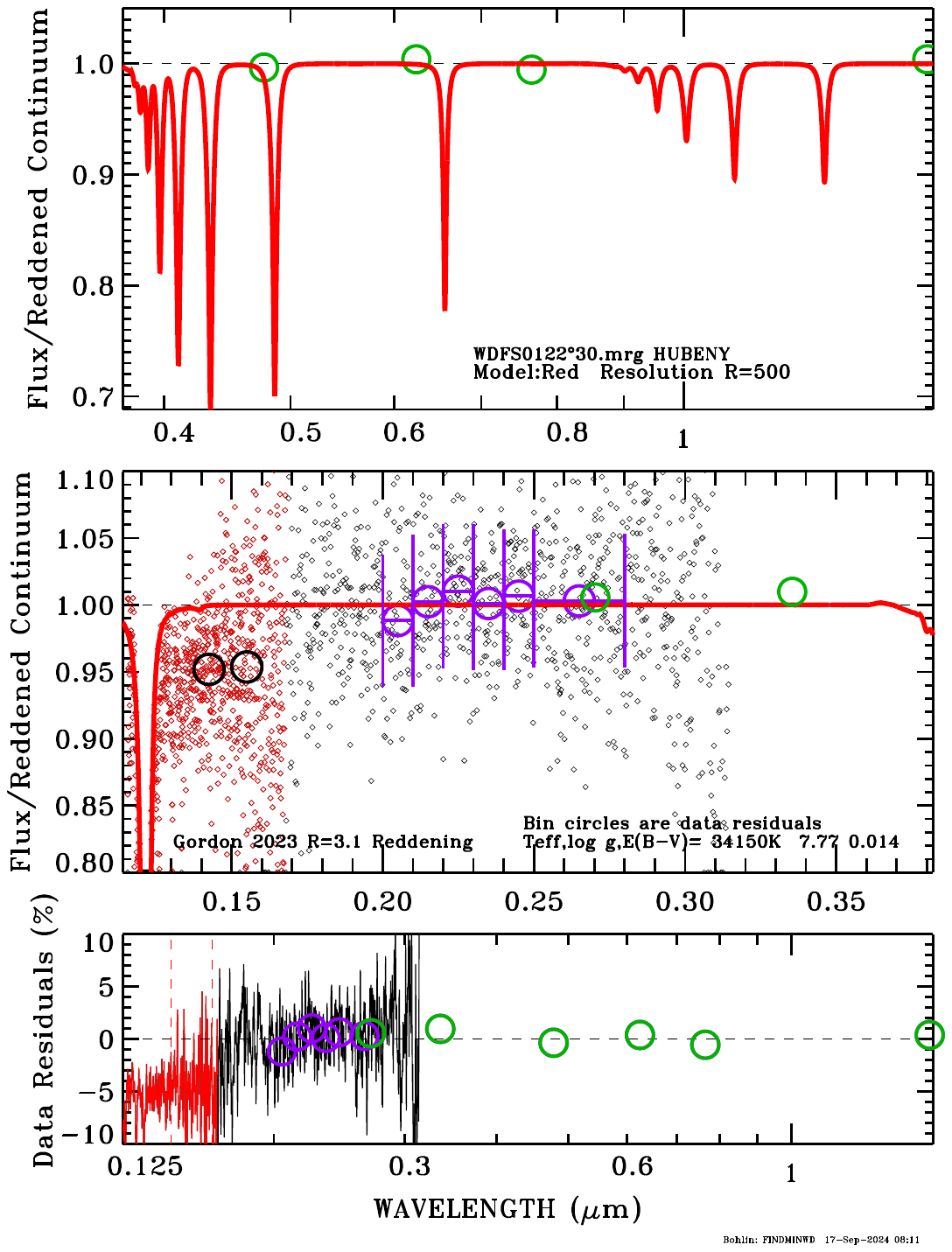}
\figsetgrpnote{Model fit and residuals for WDFS0122-30 as in
Figure~\ref{findmin1434}. The bottom panel shows the STIS spectral residuals
(purple circles) and the residuals for the WFC3 photometry (green circles).}
\figsetgrpend

\figsetgrpstart
\figsetgrpnum{figurenumber.2}
\figsetgrptitle{WDFS0248+33}
\figsetplot{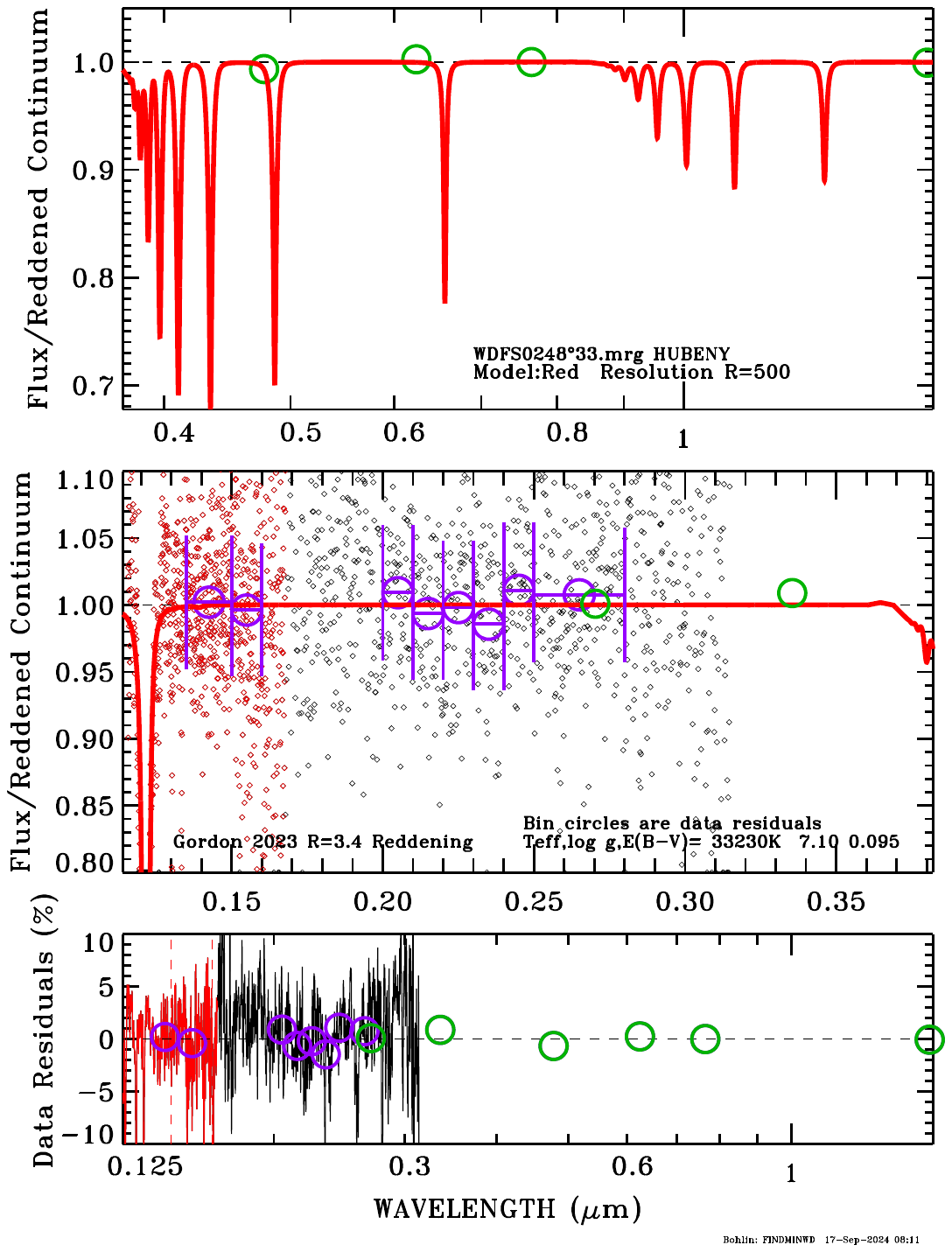}
\figsetgrpnote{Model fit and residuals for WDFS0248+33 as in
Figure~\ref{findmin1434}. The bottom panel shows the STIS spectral residuals
(purple circles) and the residuals for the WFC3 photometry (green circles)}
\figsetgrpend

\figsetgrpstart
\figsetgrpnum{figurenumber.3}
\figsetgrptitle{WDFS0458-56}
\figsetplot{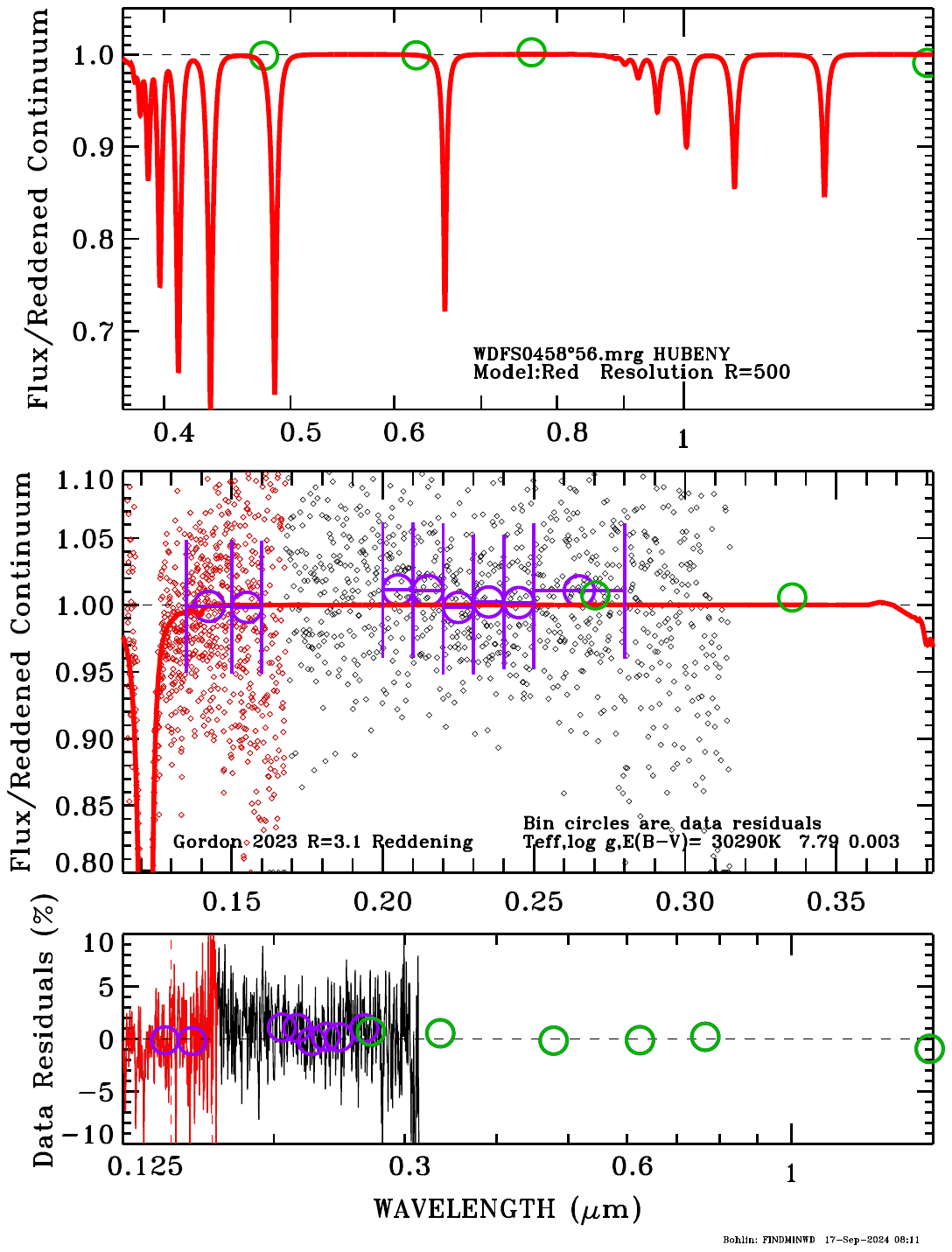}
\figsetgrpnote{Model fit and residuals for WDFS0458-56 as in
Figure~\ref{findmin1434}. The bottom panel shows the STIS spectral residuals
(purple circles) and the residuals for the WFC3 photometry (green circles)}
\figsetgrpend

\figsetgrpstart
\figsetgrpnum{figurenumber.4}
\figsetgrptitle{WDFS0639-57}
\figsetplot{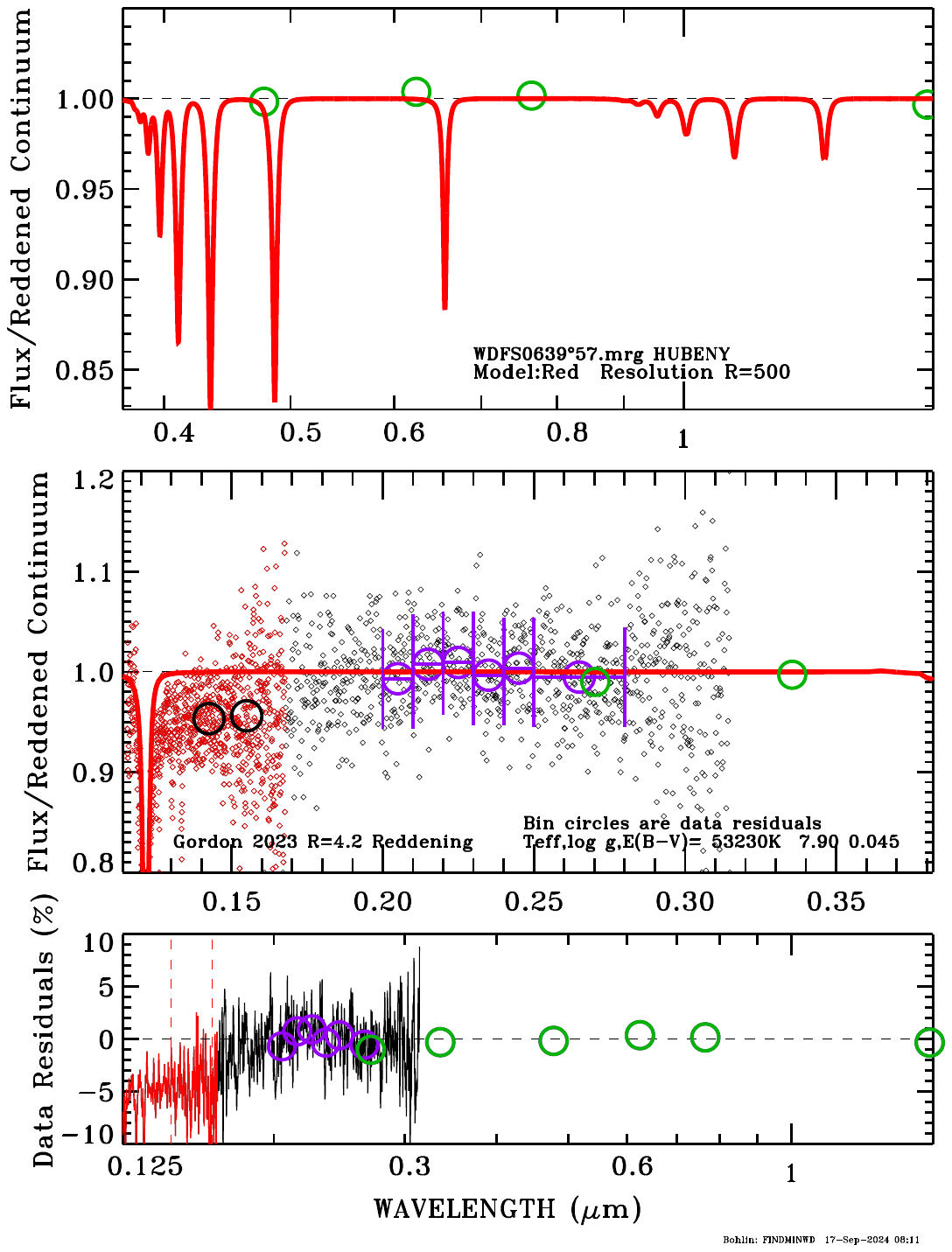}
\figsetgrpnote{Model fit and residuals for WDFS0639-57 as in
Figure~\ref{findmin1434}. The bottom panel shows the STIS spectral residuals
(purple circles) and the residuals for the WFC3 photometry (green circles)}
\figsetgrpend

\figsetgrpstart
\figsetgrpnum{figurenumber.5}
\figsetgrptitle{WDFS0956-38}
\figsetplot{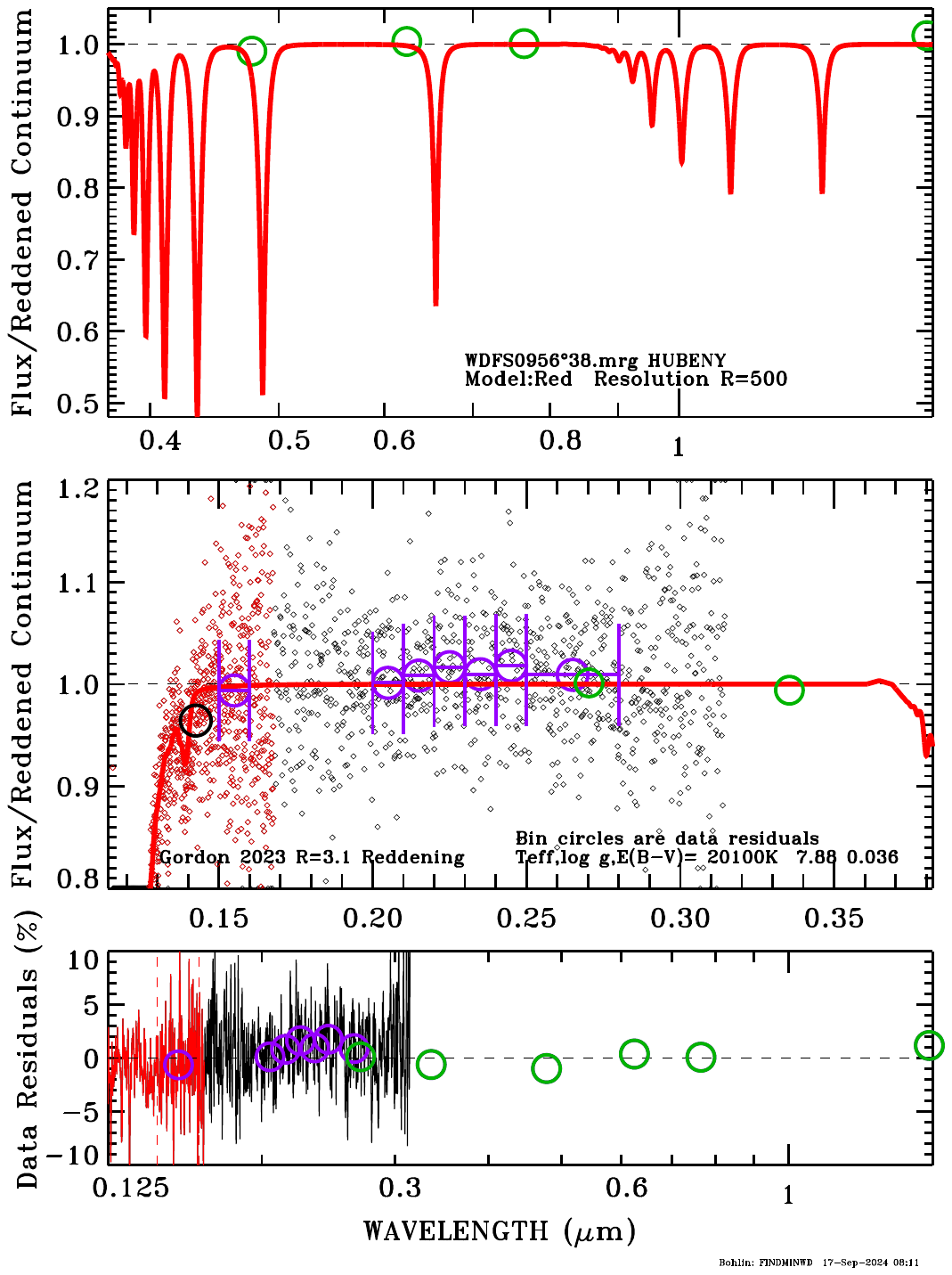}
\figsetgrpnote{Model fit and residuals for WDFS0956-38 as in
Figure~\ref{findmin1434}. The bottom panel shows the STIS spectral residuals
(purple circles) and the residuals for the WFC3 photometry (green circles)}
\figsetgrpend

\figsetgrpstart
\figsetgrpnum{figurenumber.6}
\figsetgrptitle{WDFS1055-36}
\figsetplot{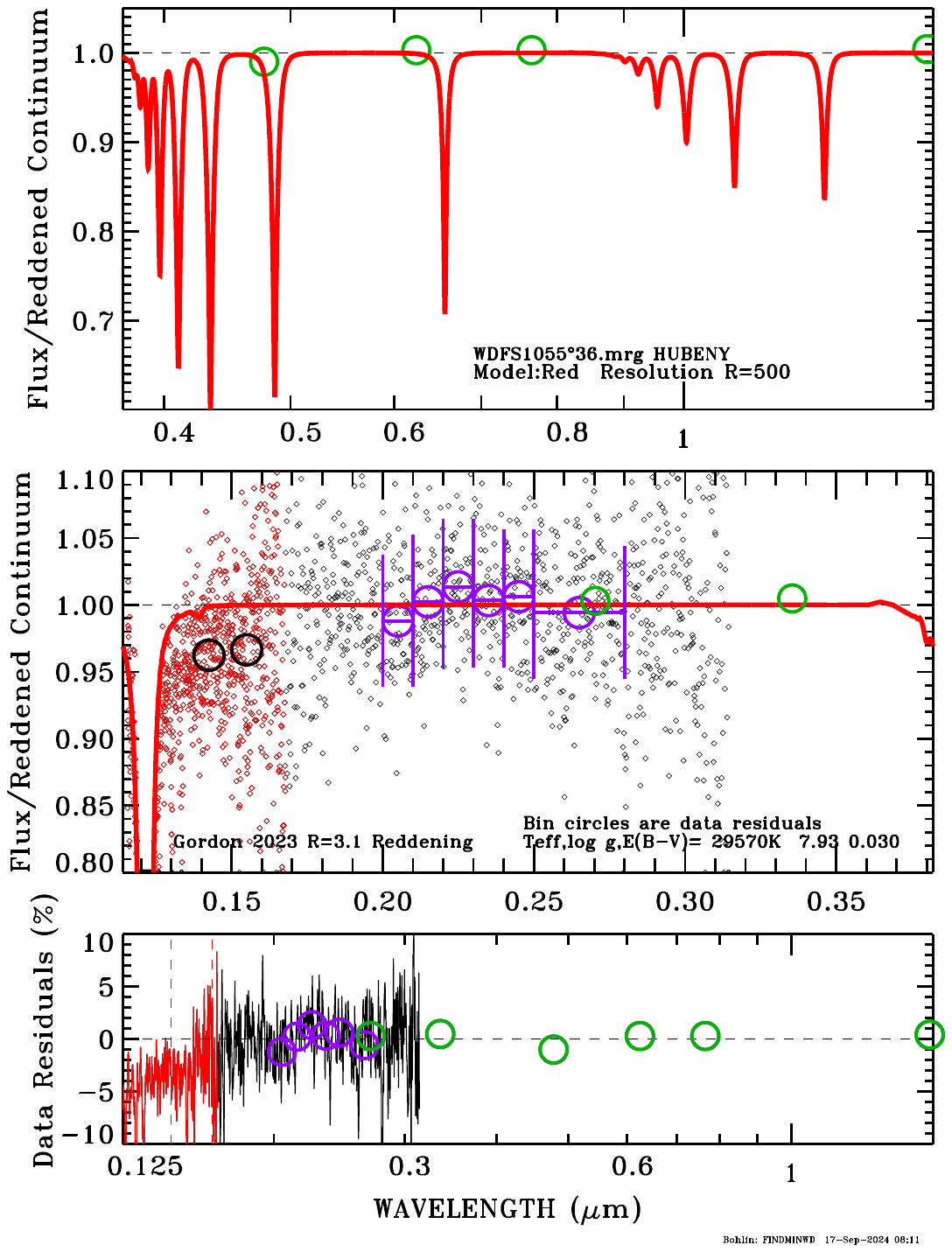}
\figsetgrpnote{Model fit and residuals for WDFS1055-36 as in
Figure~\ref{findmin1434}. The bottom panel shows the STIS spectral residuals
(purple circles) and the residuals for the WFC3 photometry (green circles)}
\figsetgrpend

\figsetgrpstart
\figsetgrpnum{figurenumber.7}
\figsetgrptitle{WDFS1110-17}
\figsetplot{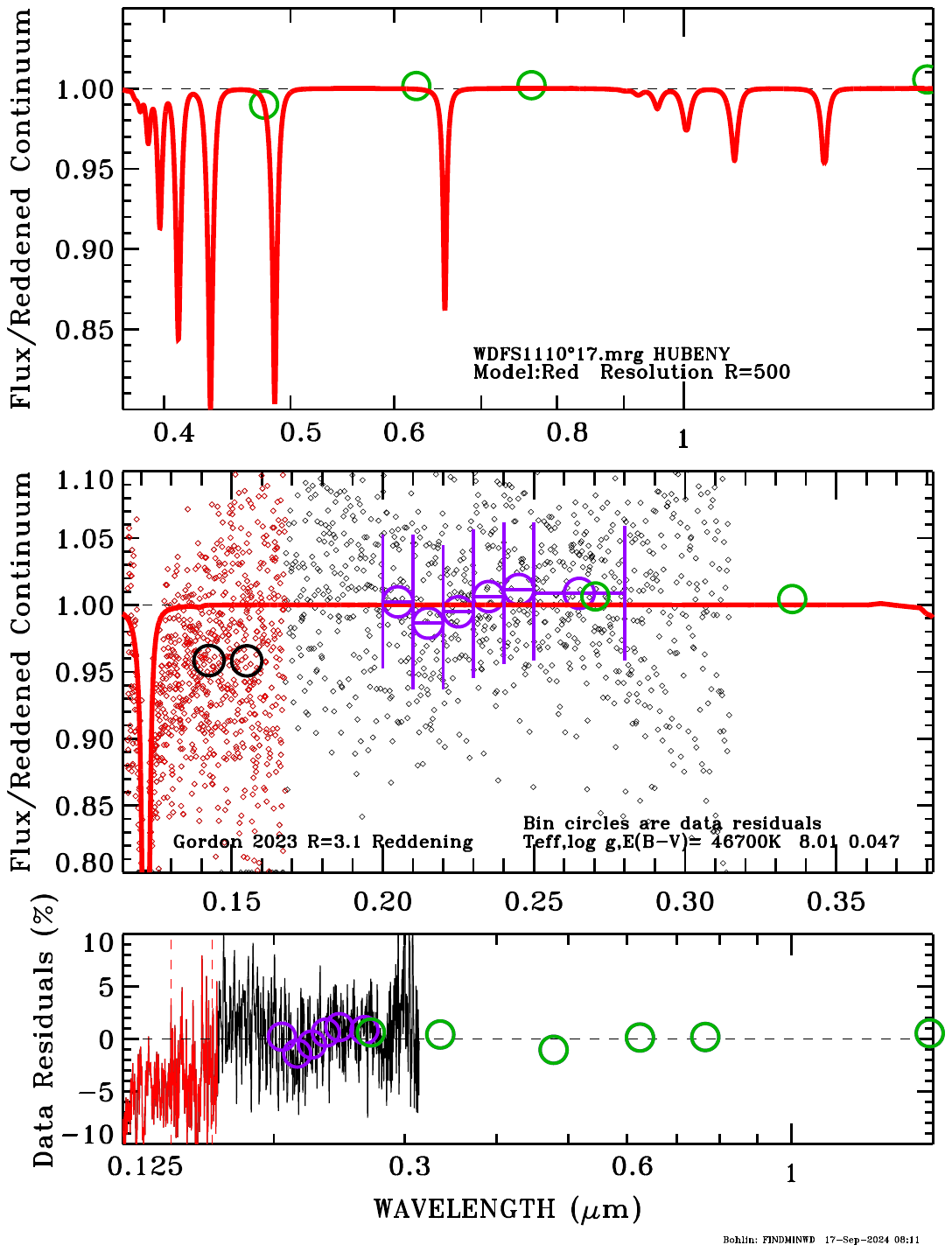}
\figsetgrpnote{Model fit and residuals for WDFS1110-17 as in
Figure~\ref{findmin1434}. The bottom panel shows the STIS spectral residuals
(purple circles) and the residuals for the WFC3 photometry (green circles)}
\figsetgrpend

\figsetgrpstart
\figsetgrpnum{figurenumber.8}
\figsetgrptitle{WDFS1206-27}
\figsetplot{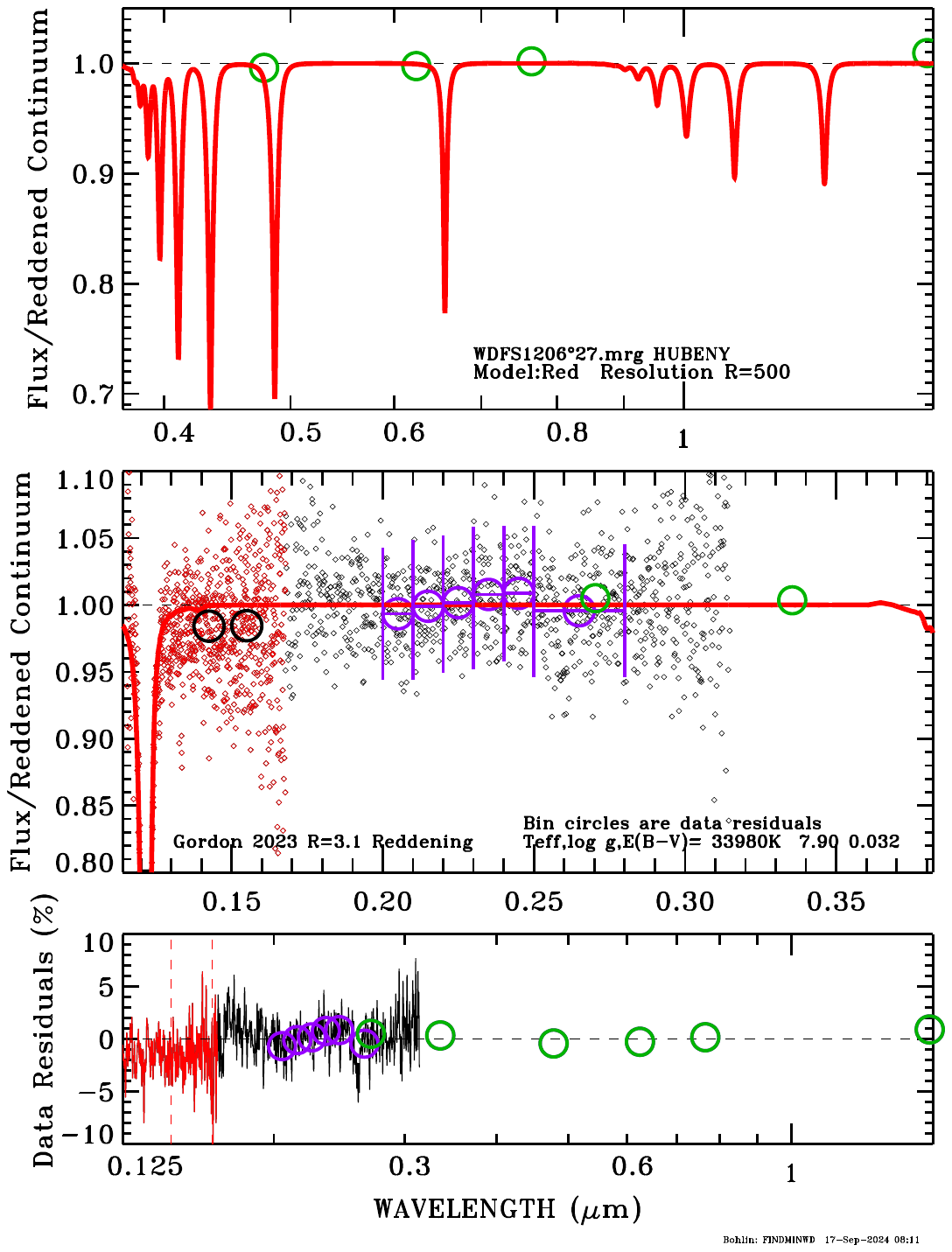}
\figsetgrpnote{Model fit and residuals for WDFS1206-27 as in
Figure~\ref{findmin1434}. The bottom panel shows the STIS spectral residuals
(purple circles) and the residuals for the WFC3 photometry (green circles)}
\figsetgrpend

\figsetgrpstart
\figsetgrpnum{figurenumber.9}
\figsetgrptitle{WDFS1214+45}
\figsetplot{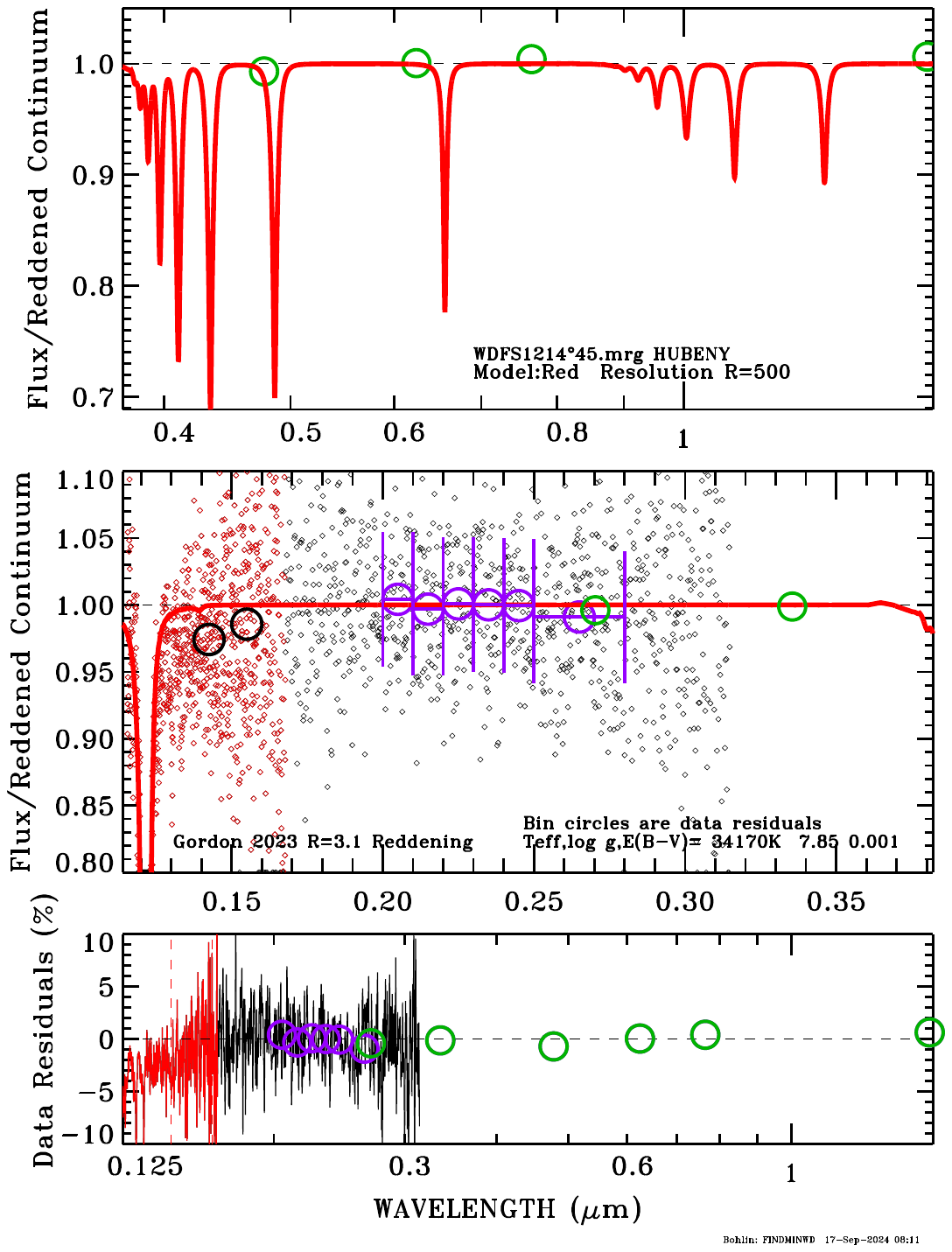}
\figsetgrpnote{Model fit and residuals for WDFS1214+45 as in
Figure~\ref{findmin1434}. The bottom panel shows the STIS spectral residuals
(purple circles) and the residuals for the WFC3 photometry (green circles)}
\figsetgrpend

\figsetgrpstart
\figsetgrpnum{figurenumber.10}
\figsetgrptitle{WDFS1302+10}
\figsetplot{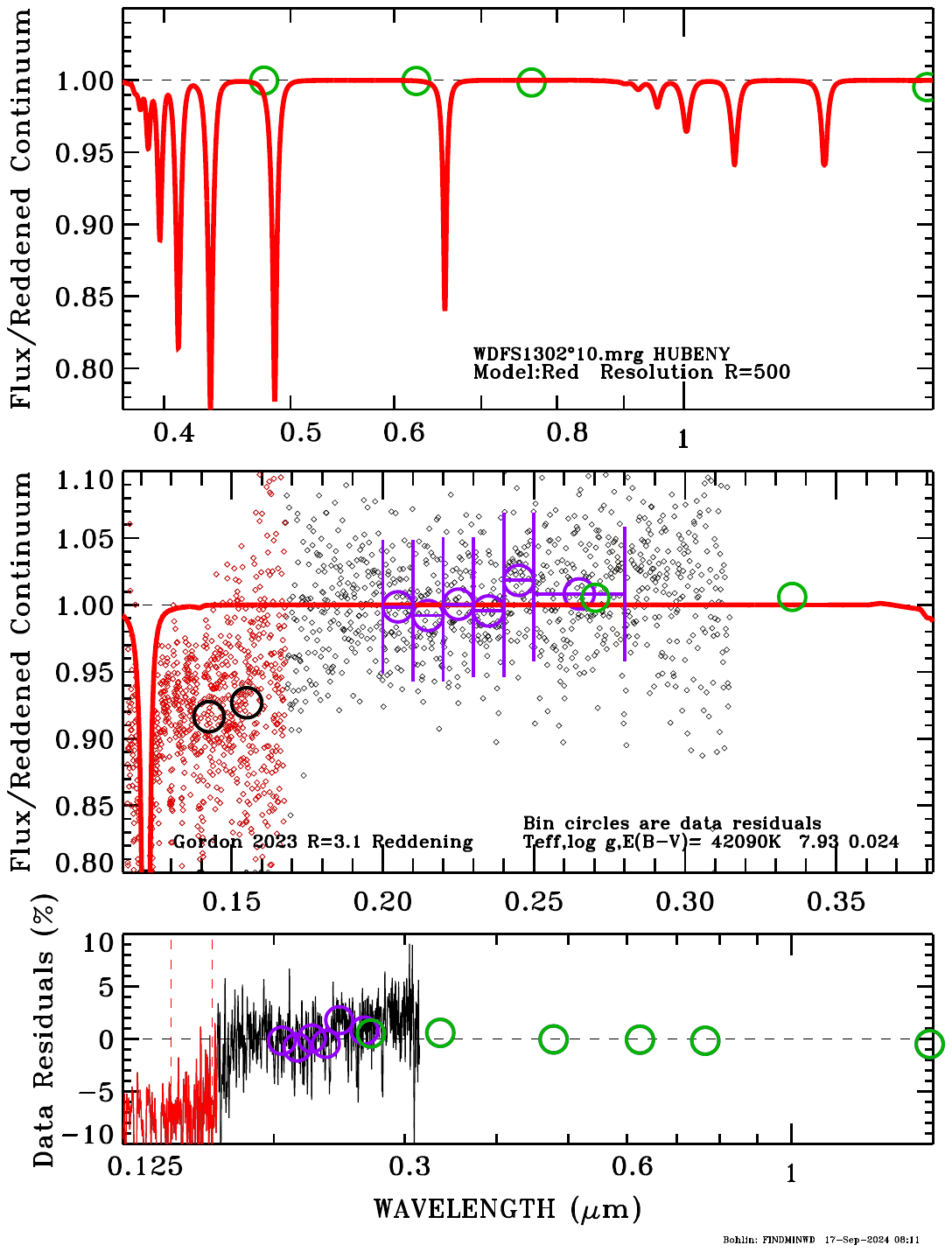}
\figsetgrpnote{Model fit and residuals for WDFS1302+10 as in
Figure~\ref{findmin1434}. The bottom panel shows the STIS spectral residuals
(purple circles) and the residuals for the WFC3 photometry (green circles)}
\figsetgrpend

\figsetgrpstart
\figsetgrpnum{figurenumber.12}
\figsetgrptitle{WDFS1514+00}
\figsetplot{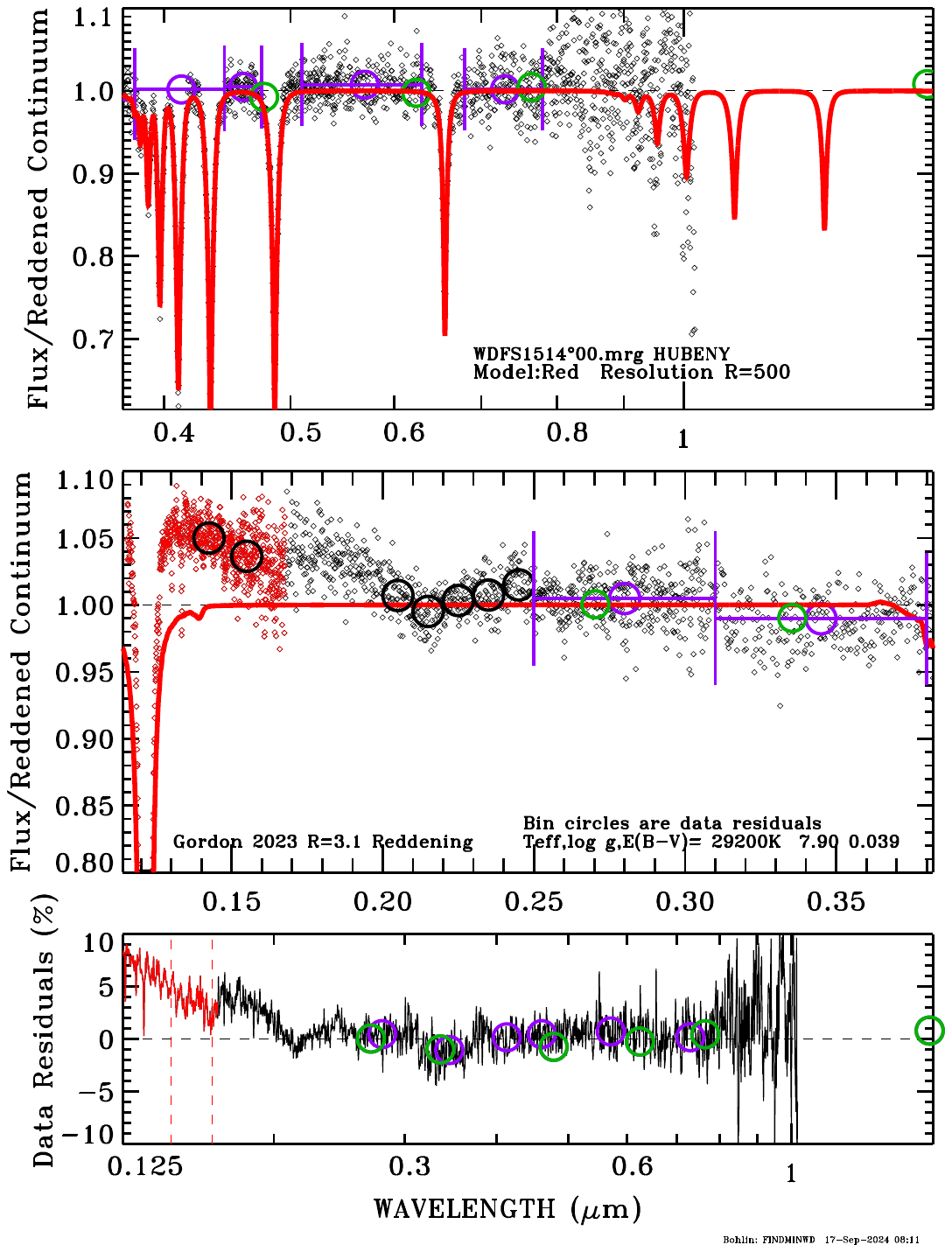}
\figsetgrpnote{Model fit and residuals for WDFS1514+00 as in
Figure~\ref{findmin1434}. The bottom panel shows the STIS spectral residuals
(purple circles) and the residuals for the WFC3 photometry (green circles)}
\figsetgrpend

\figsetgrpstart
\figsetgrpnum{figurenumber.13}
\figsetgrptitle{WDFS1535-77}
\figsetplot{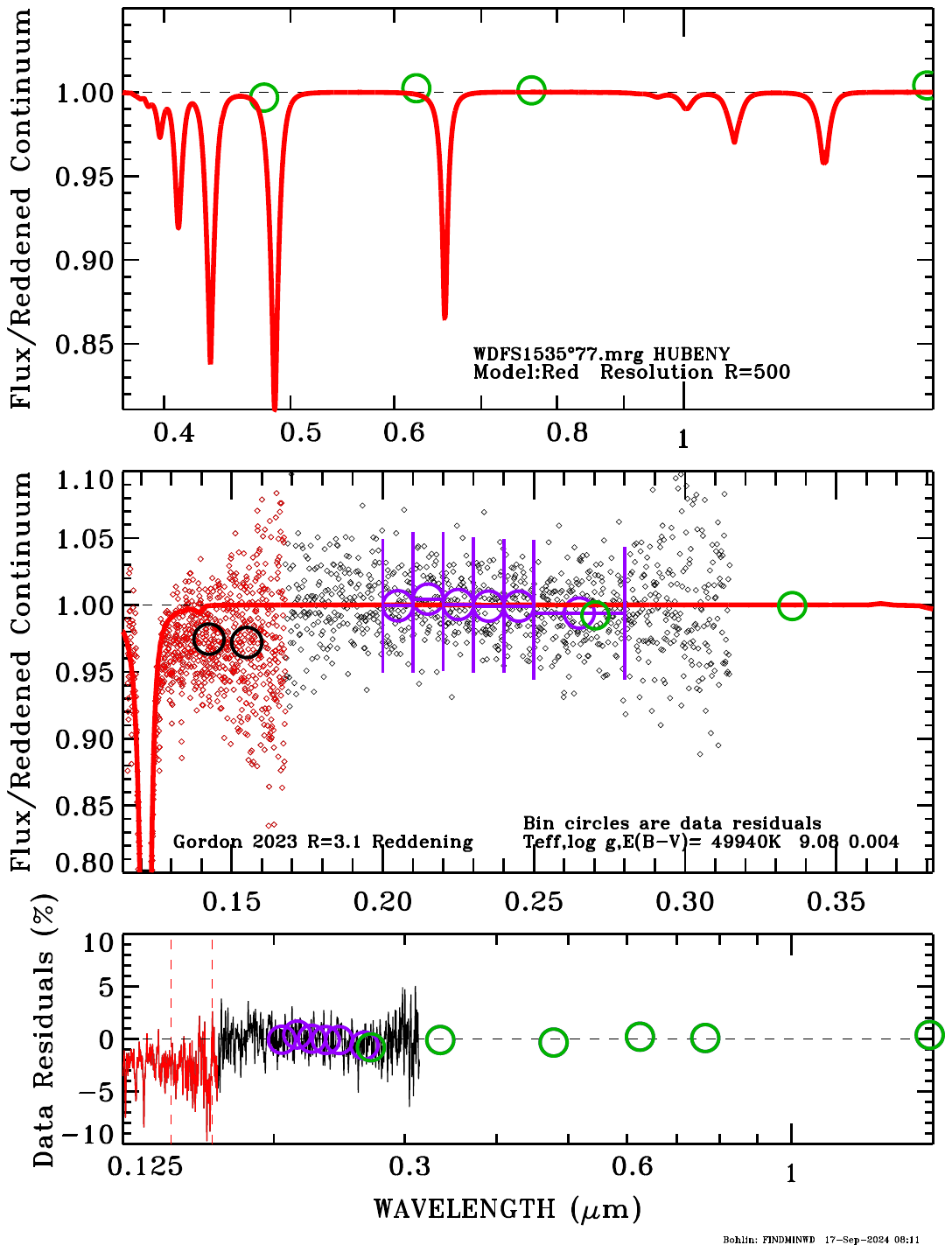}
\figsetgrpnote{Model fit and residuals for WDFS1535-77 as in
Figure~\ref{findmin1434}. The bottom panel shows the STIS spectral residuals
(purple circles) and the residuals for the WFC3 photometry (green circles)}
\figsetgrpend

\figsetgrpstart
\figsetgrpnum{figurenumber.14}
\figsetgrptitle{WDFS1557+55}
\figsetplot{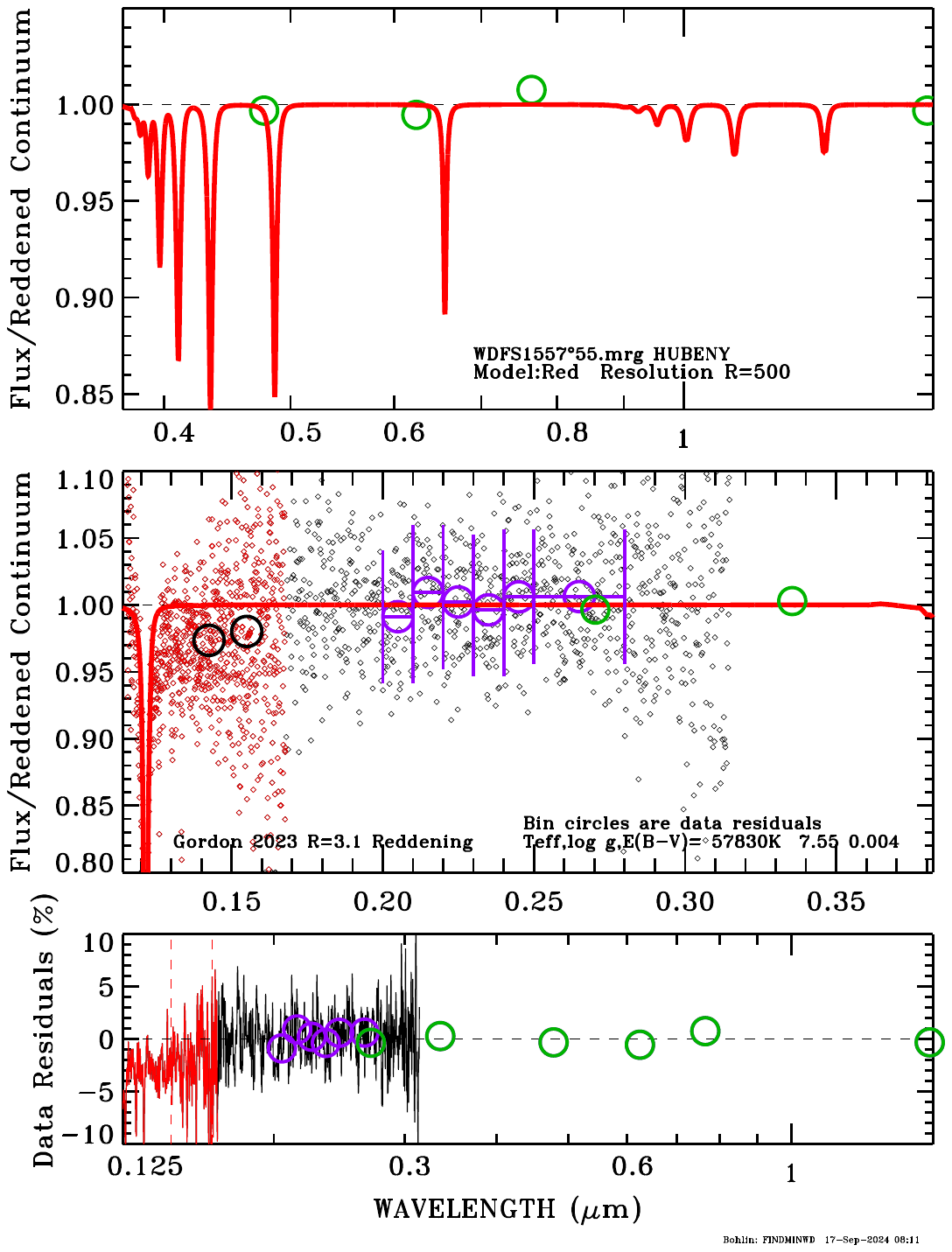}
\figsetgrpnote{Model fit and residuals for WDFS1557+55 as in
Figure~\ref{findmin1434}. The bottom panel shows the STIS spectral residuals
(purple circles) and the residuals for the WFC3 photometry (green circles)}
\figsetgrpend

\figsetgrpstart
\figsetgrpnum{figurenumber.15}
\figsetgrptitle{WDFS1814+78}
\figsetplot{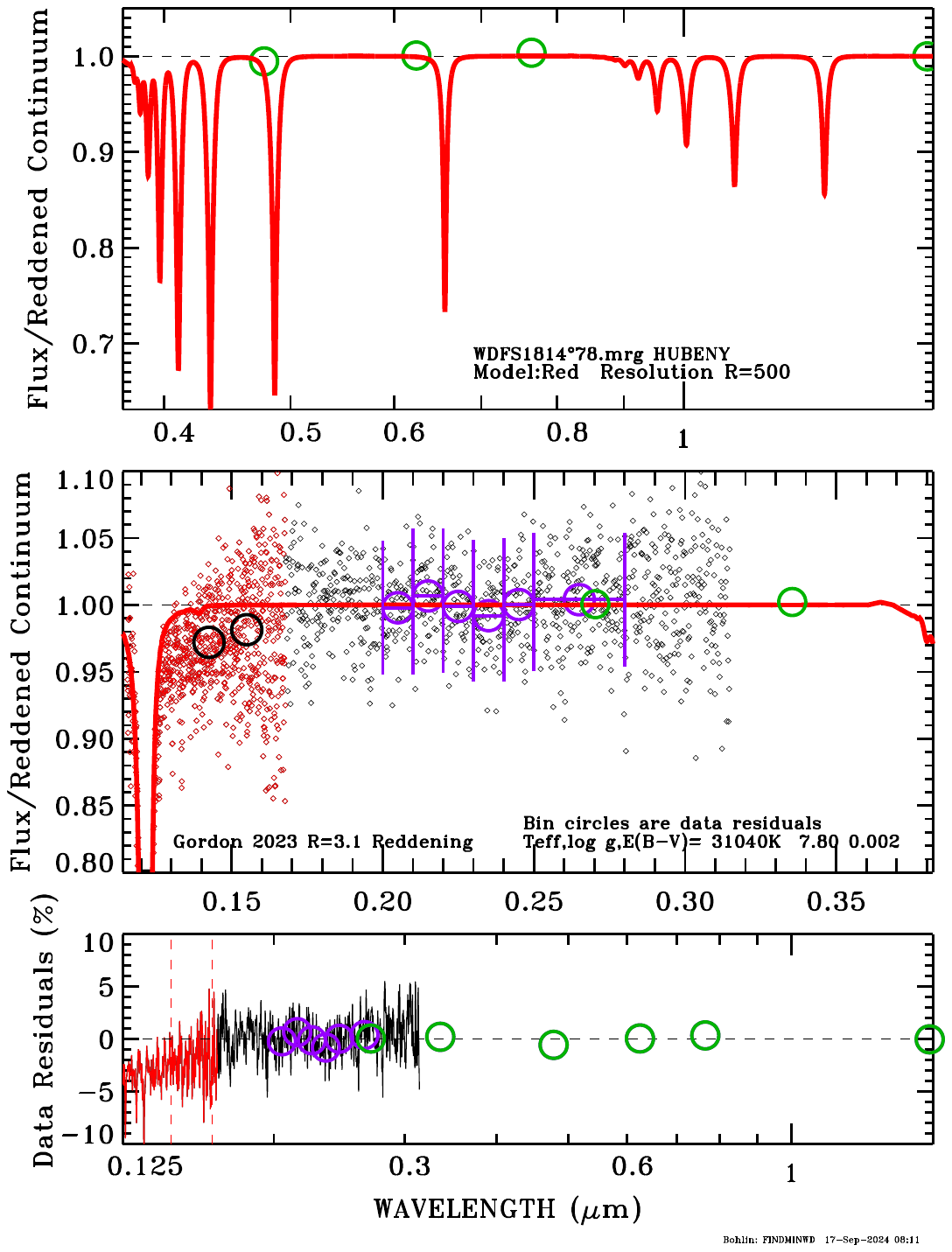}
\figsetgrpnote{Model fit and residuals for WDFS1814+78 as in
Figure~\ref{findmin1434}. The bottom panel shows the STIS spectral residuals
(purple circles) and the residuals for the WFC3 photometry (green circles)}
\figsetgrpend

\figsetgrpstart
\figsetgrpnum{figurenumber.16}
\figsetgrptitle{WDFS1837-70}
\figsetplot{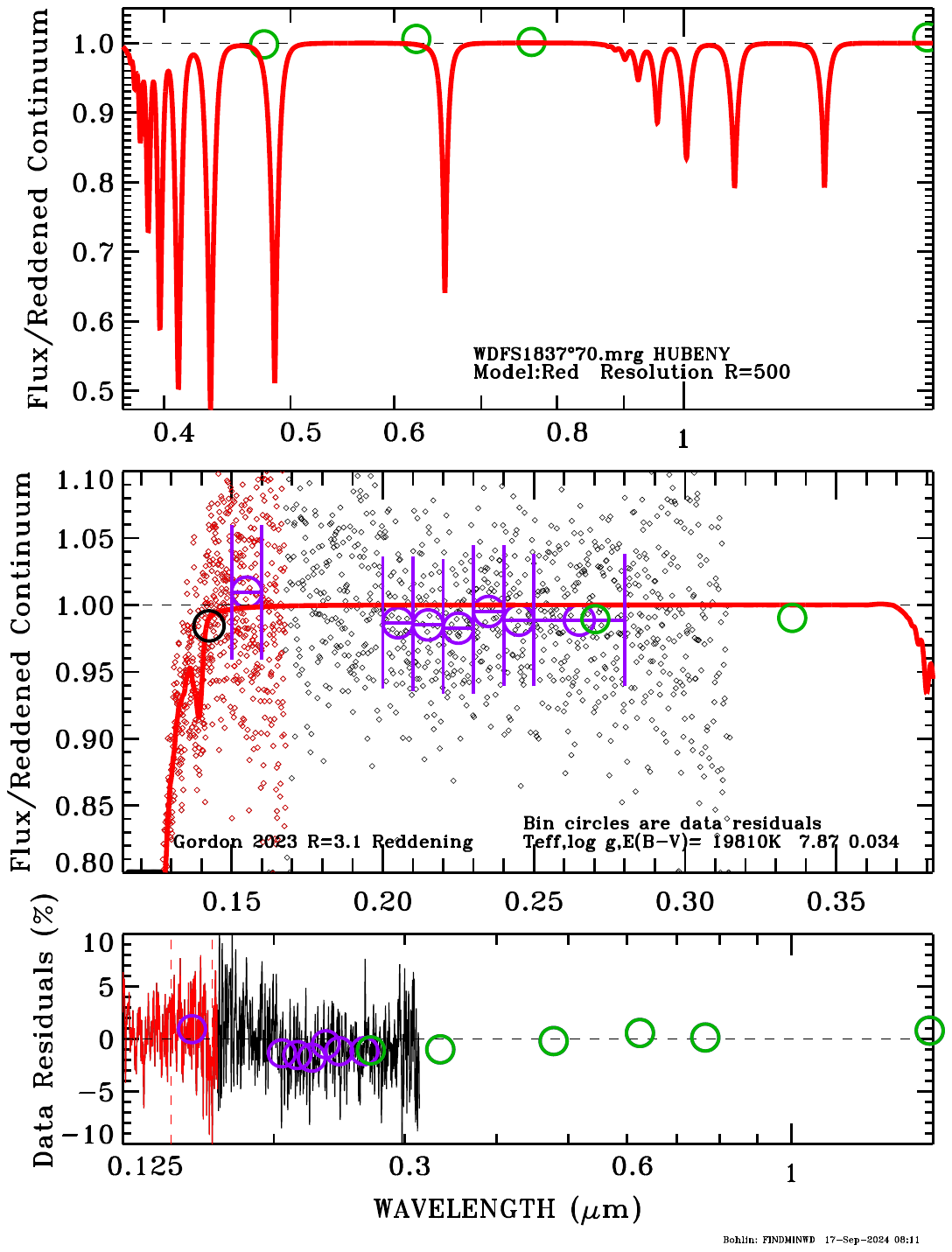}
\figsetgrpnote{Model fit and residuals for WDFS1837-70 as in
Figure~\ref{findmin1434}. The bottom panel shows the STIS spectral residuals
(purple circles) and the residuals for the WFC3 photometry (green circles)}
\figsetgrpend

\figsetgrpstart
\figsetgrpnum{figurenumber.17}
\figsetgrptitle{WDFS1930-52}
\figsetplot{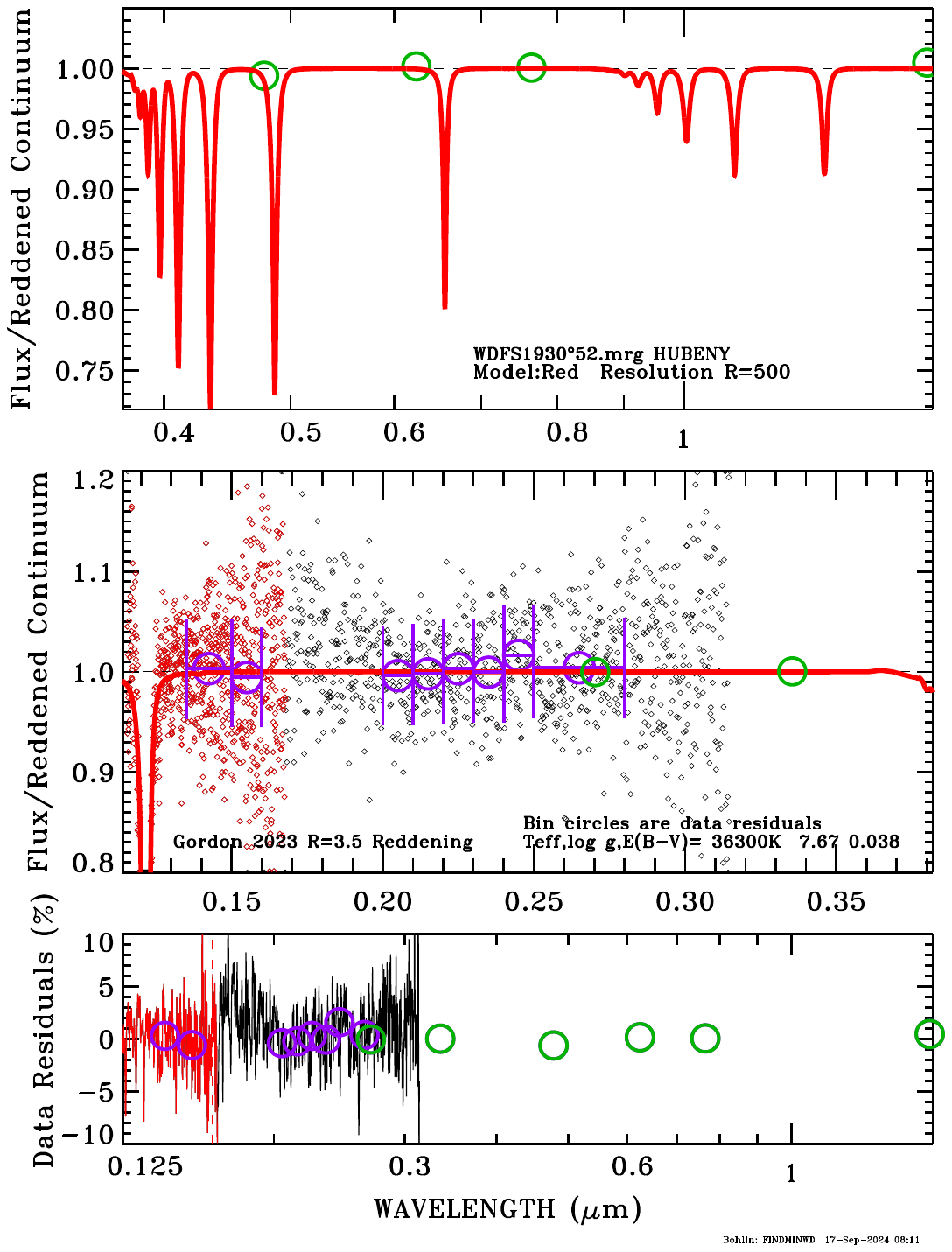}
\figsetgrpnote{Model fit and residuals for WDFS1930-52 as in
Figure~\ref{findmin1434}. The bottom panel shows the STIS spectral residuals
(purple circles) and the residuals for the WFC3 photometry (green circles)}
\figsetgrpend

\figsetgrpstart
\figsetgrpnum{figurenumber.18}
\figsetgrptitle{WDFS2317-29}
\figsetplot{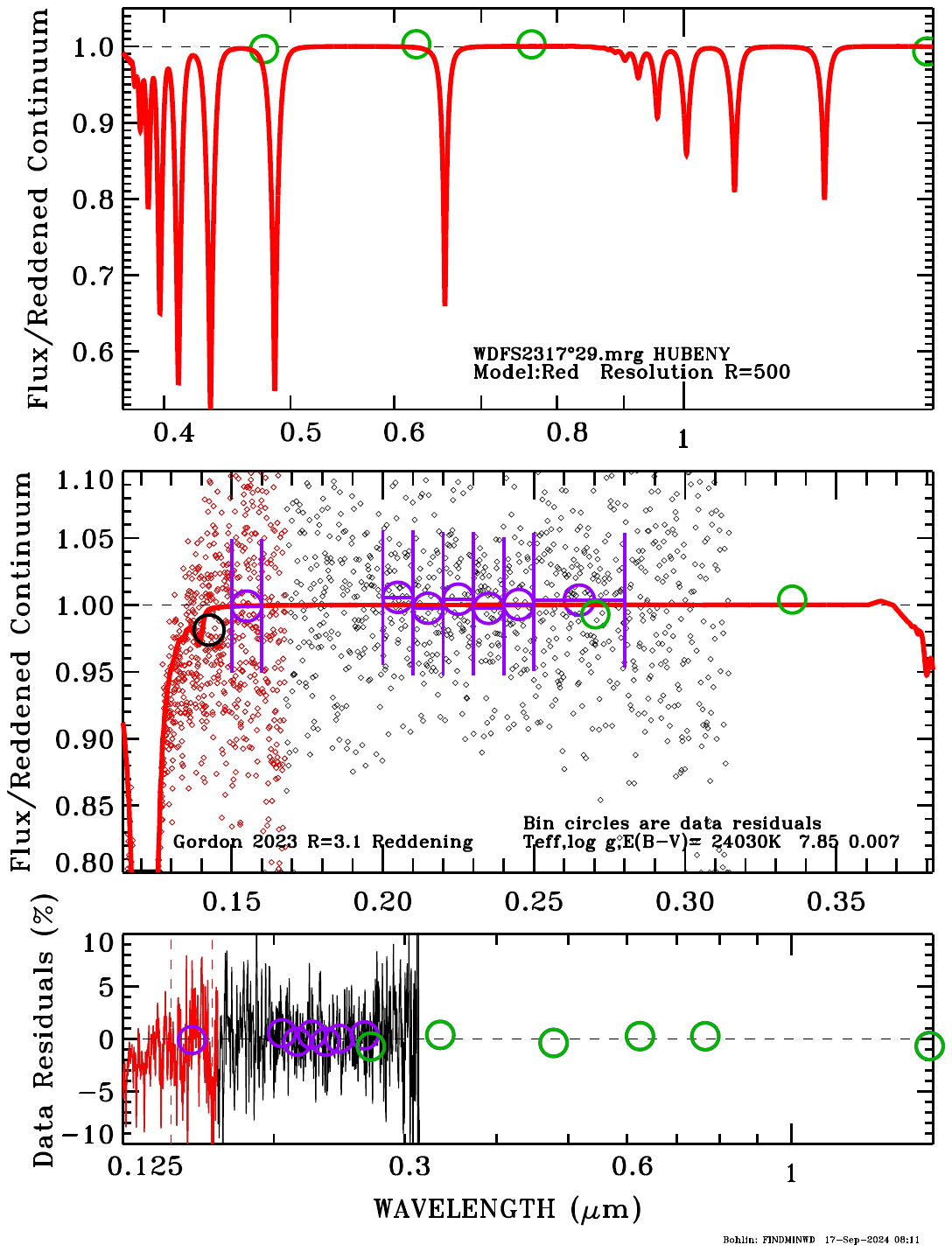}
\figsetgrpnote{Model fit and residuals for WDFS2317-29 as in
Figure~\ref{findmin1434}. The bottom panel shows the STIS spectral residuals
(purple circles) and the residuals for the WFC3 photometry (green circles)}
\figsetgrpend
\figsetend


\subsection{Statistics of the Fits}				

Table~\ref{table:stats} contains the average offsets and their uncertainties of
the deviation of the WFC3 photometry from the synthetic photometry for the 19
fitted models. The uncertainties, i.e. errors-in-the-mean, for the offsets are
the rms values divided by $\sqrt19$. These results are consistent with the
\citet{axelrod2023} corrections to the original WFC3 photometry, except for
slightly larger uncertainties.

\subsection{SED Uncertainties}				

The Hubeny model atmosphere grid is compared to the similar Rauch grid
\citep{bohlin2020} to verify their near equivalence and provide a lower limit to
uncertainty of our final Hubeny SEDs.  The Rauch \textsc{tmap} models are nearly
identical to the Hubeny \textsc{tlusty} models, except for the quasi-molecular
hydrogen FUV absorption features in the \textsc{tlusty} models
\citep{bohlin2020}. Model fits with the \textsc{tlusty} or \textsc{tmap} grids
are generally within 1\% of agreement. An outlier is WDFS0248+33, where the
Rauch fit is 2\% brighter than Hubeny at 30~\micron. In the range of STIS and
WFC3 observational constraints, the Table~\ref{table:modpar} model fits and the
\citet{axelrod2023} SEDs agree to 1\%; but in the mid-IR, a minimum
uncertainty is the 2\% set by the worst agreement between the fits for the two 
independent grids. However, the actual errors in the IR can be larger, if
noise or some systematic affects both the \textsc{tlusty} and \textsc{tmap}
fits similarly.

The best measure of uncertainty in the important CALSPEC extrapolations past the
long wavelength constraint of F160W with a pivot wavelength of 15369~\AA\ is
defined by how well the current brighter standards work for JWST preliminary
flux calibrations. The presently limited set of CALSPEC SEDs are derived from
fitting HST flux measures with the same $\chi^2$ technique used here. These
preliminary JWST flux calibrations are generally consistent in the
2--30~\micron\ range to better than $\approx$3\% (Gordon et al. 2024, submitted).
At shorter wavelengths in the 0.27 to 1.6~\micron\ range with WFC3 photometry,
an uncertainty of $\approx$1\% is demonstrated by that level of agreement with
the seminal results of \citet{axelrod2023}.

\begin{deluxetable}{lrc} 			
\tablewidth{0pt}
\tablecaption{\label{table:stats} \mbox{Average WFC3 Offsets from the Model
and}
\mbox{their Uncertainties in Mag Units}}
\tablehead{
\colhead{Filter} &\colhead{Avg} &\colhead{Uncert.}}
\startdata
F275W  &+0.001  &0.006 \\
F336W  &-0.002  &0.006 \\
F475W  &+0.006  &0.004 \\
F625W  &-0.001  &0.003 \\
F775W  &-0.002  &0.003 \\
F160W  &-0.003  &0.006 \\
\enddata
\end{deluxetable}

\section{Discussion}			

Our model fits to the HST STIS spectroscopic plus WFC3 photometric
observations provide continuous wavelength coverage from 900~\AA\ to 30~\micron\
at a resolution R=5,000. The mid-IR
flux distributions to 30~\micron\ are essential for the flux calibration of the
JWST instruments. Consistency of the JWST instrumental calibrations from our WDs
in comparison to results for A and G type stars would validate our WD model
extrapolations into the mid-IR.

While our results will be refined by on-going studies, the present results build
on the comprehensive \citet{axelrod2023} work and are adequate for an initial
delivery to 
CALSPEC\footnote{http://www.stsci.edu/hst/instrumentation/reference-data-for-calibration-and-tools/astronomical-catalogs/calspec},
in order to provide our all-sky network of faint flux standards for calibration
of JWST and other large-aperture facilities such as Euclid, the Nancy Grace
Roman Telescope, and the Vera C. Rubin Observatory. With the new STIS data, NLTE
models with wavelength coverage to 30~\micron, and  SEDs on the current HST
CALSPEC flux scale, our updated SEDs are immediately useful. Future revisions
will be posted, as is the usual custom for CALSPEC quality SEDs, which track
on-going improvements to the data reduction and to the model fitting techniques.

The HST data presented in this article were obtained from the Mikulski Archive
for Space Telescopes (MAST) at the Space Telescope Science Institute. The
specific observations analyzed can be accessed via
\dataset[10.17909/bwd3-9z83]{https://doi.org/10.17909/bwd3-9z83}.

\section*{Acknowledgements}

Support for this work was provided by NASA through the Space Telescope Science
Institute, which is operated by AURA, Inc., under NASA contract NAS5-26555.

\textbf{ORCID IDs}

Ralph C. Bohlin https://orcid.org/0000-0001-9806-0551 \newline
Abhijit Saha    https://orcid.org/0000-0002-6839-4881 \newline
Gautham Narayan https://orcid.org/0000-0001-6022-0484 \newline
Annalisa Calamida https://orcid.org/0000-0002-0882-7702 \newline
Susana Deustua  https://orcid.org/0000-0003-2823-360X \newline
Karl D. Gordon  https://orcid.org/0000-0001-5340-6774 \newline
Jay B. Holberg  https://orcid.org/0000-0003-3082-0774 \newline
Ivan Hubenby    https://orcid.org/0000-0001-8816-236X \newline
Thomas Matheson https://orcid.org/0000-0001-6685-0479 \newline
Armin Rest	https://orcid.org/0000-0002-4410-5387 \newline
\bibliographystyle{/Users/bohlin/pub/apj.bst}
\bibliography{/Users/bohlin/pub/paper-bibliog.bib}
\end{document}